\documentclass{emulateapj}
\usepackage{graphicx}
\usepackage{epstopdf}
\usepackage{color}

\newcommand{\Teff}{\mbox{$T_\mathrm{eff}$}}
\newcommand{\Mjup}{\mbox{$M_\mathrm{Jup}$}}
\newcommand{\Msun}{\mbox{$M_{\odot}$}}

\begin{document}
\title{Planets Around Low-Mass Stars (PALMS). \\ III.  A Young Dusty L Dwarf Companion at the Deuterium-Burning Limit$^*$$^{\dag}$}
\author{Brendan P. Bowler,\altaffilmark{1} 
Michael C. Liu,\altaffilmark{1,2} 
Evgenya L. Shkolnik,\altaffilmark{3}
Trent J. Dupuy\altaffilmark{4, 5}
\\ }
\email{bpbowler@ifa.hawaii.edu}

\altaffiltext{1}{Institute for Astronomy, University of Hawai`i; 2680 Woodlawn Drive, Honolulu, HI 96822, USA}
\altaffiltext{2}{Visiting Astronomer at the Infrared Telescope Facility, which is operated by the University of Hawai`i 
under Cooperative Agreement no. NNX-08AE38A with the National Aeronautics and Space Administration, 
Science Mission Directorate, Planetary Astronomy Program.}
\altaffiltext{3}{Lowell Observatory, 1400 W. Mars Hill Road, Flagstaff, AZ 86001}
\altaffiltext{4}{Harvard-Smithsonian Center for Astrophysics, 60 Garden Street, Cambridge, MA 02138}
\altaffiltext{5}{Hubble Fellow}
\altaffiltext{*}{Some of the data presented herein were obtained at the W.M. Keck Observatory, which is operated as a 
scientific partnership among the California Institute of Technology, the University of California and the National 
Aeronautics and Space Administration. The Observatory was made possible by the generous financial support of the W.M. Keck Foundation.}
\altaffiltext{$\dag$}{Based in part on data collected at Subaru Telescope, which is operated by the National Astronomical Observatory of Japan.}

\submitted{ApJ, Accepted (July 7 2013)}

\begin{abstract}

We report the discovery of an L-type companion to the young M3.5V star 
2MASS~J01225093--2439505 at a projected separation of 1.45$''$ ($\approx$52~AU)
as part of our adaptive optics
imaging search for extrasolar giant planets around young low-mass stars.  
2MASS~0122--2439~B has very red near-infrared colors 
similar to the HR 8799 planets and
the reddest known young/dusty L dwarfs in the field.
Moderate-resolution ($R$$\approx$3800) 1.5--2.4~$\mu$m spectroscopy 
reveals a near-infrared spectral type of L4--L6 and an angular $H$-band shape,
confirming its cool temperature and young age.
The kinematics of 2MASS~0122--2439~AB are marginally consistent with 
members of the $\sim$120~Myr AB~Dor young moving group based on
the photometric distance to the primary (36~$\pm$~4~pc) and our radial velocity measurement 
of 2MASS~0122--2439~A from Keck/HIRES. 
We adopt the AB~Dor group age for the system, 
but the high energy emission, lack of \ion{Li}{1} $\lambda$6707 
absorption, and spectral shape of 2MASS~0122--2439~B suggest a range of $\sim$10--120~Myr is possible.
The age and luminosity of 2MASS~0122--2439~B fall in a strip
where ``hot-start''  evolutionary model mass tracks overlap as a result of deuterium burning.
Several known substellar companions also fall in this region
(2MASS~J0103--5515~ABb, AB~Pic~b, $\kappa$~And~b, G196-3~B, SDSS~2249+0044~B, 
LP~261-75~B, HD 203030 B, and HN Peg B),
but their dual-valued mass predictions have largely been unrecognized.
The implied mass of 2MASS~0122--2439~B is $\approx$12--13~\Mjup \
or $\approx$22--27~\Mjup \ if it is an AB~Dor member, or possibly as low as 
11~\Mjup \ if
the wider age range is adopted.
Evolutionary models predict an 
effective temperature for 2MASS~0122--2439~B that corresponds to spectral types near the 
L/T transition ($\approx$1300--1500~K) for field objects.
However, we find a mid-L near-infrared spectral type, 
indicating that 2MASS~0122--2439~B represents another case of 
photospheric dust being retained to cooler temperatures 
at low surface gravities, as seen in the spectra of young (8--30~Myr) planetary companions.
Altogether, the low mass, low temperature, and red colors of 2MASS~0122--2439~B 
make it a bridge 
between warm planets like $\beta$~Pic~b and cool, very dusty ones like HR~8799~bcde.

\end{abstract}

\section{Introduction}{\label{sec:intro}}

Adaptive optics imaging 
is a powerful tool to explore the architecture and 
demographics of planetary systems beyond $\sim$10~AU.
A handful of companions below the 
deuterium-burning limit ($\sim$13~\Mjup; \citealt{Spiegel:2011p22104}) have been found,
but the origin of these planetary-mass companions is ambiguous since 
there is growing evidence that 
objects in the 5--15~\Mjup \ range can form in multiple ways.
The $\beta$~Pic and HR~8799 planets are nested in debris disks and probably 
originated in massive protoplanetary disks 
(\citealt{Marois:2008p18841}; \citealt{Marois:2010p21591}; \citealt{Lagrange:2010p21645}), 
although their exact mode of formation is still unclear 
(\citealt{DodsonRobinson:2009p19734}; \citealt{Nero:2009p20063}; \citealt{Meru:2010p21335}).
Planetary-mass companions have also been found in distinctly ``non-planetary'' environments ---
both orbiting stars at ultra-wide separations (\citealt{Lafreniere:2008p14057}; \citealt{Ireland:2011p21592}; \citealt{Luhman:2011p22766}), 
and brown dwarfs with large mass ratios (\citealt{Chauvin:2004p19400}; \citealt{Todorov:2010p20562}; \citealt{Liu:2011p22852}) ---
pointing to an origin more consistent with molecular cloud fragmentation than formation in a disk
(\citealt{Lodato:2005p20130}; \citealt{Bate:2012p24259}; \citealt{Vorobyov:2013p24492}).
These multiple possibilities make the formation scenario of any individual system difficult to deduce
if devoid of environmental clues (\citealt{Bowler:2011p23014}; \citealt{Bailey:2013p24473}).

Direct imaging also enables detailed studies of planetary atmospheres through spectroscopy.
Follow-up observations of the young gas giants orbiting HR~8799
 (\citealt{Bowler:2010p21344}; \citealt{Barman:2011p22098}; \citealt{Konopacky:2013p24810}; \citealt{Oppenheimer:2013p24690})
and the planetary-mass companion 2MASS~1207--3932~b 
(\citealt{Skemer:2011p22100}; \citealt{Barman:2011p22429})
confirmed the surprising gravity-dependence of the L-T transition for young L dwarfs
(e.g., \citealt{Metchev:2006p10342}; \citealt{Stephens:2009p19484}).
In cool, low-gravity atmospheres, 
photospheric dust appears to be retained to lower effective temperatures 
compared to old (high gravity) brown dwarfs in the field.
\citet{Marley:2012p24017} provide an intuitive framework for this phenomenon
in which the gravity-dependence of cloud particle size and cloud base pressure
conspire to produce similar column optical depths in cool, low-gravity
photospheres and slightly warmer, high-gravity conditions.
This results in spectral types that appear earlier than expected based on  
evolutionary model-predicted temperatures and spectral type-temperature relations for field objects.
However, a detailed understanding of this process is
hampered by the small number of young L and T dwarfs known.

We are conducting a deep adaptive optics imaging survey of young (10--300~Myr) low-mass
(0.2--0.6~\Msun) stars to study the outer architectures of M dwarf planetary systems 
and to identify giant planets for spectral characterization.
In \citet{Bowler:2012p23851} and \citet{Bowler:2012p23980} we presented the discovery of two 
young ($\sim$120--300~Myr) substellar companions  
with masses of 32~$\pm$~6~\Mjup \ and 46~$\pm$~16~\Mjup \ at separations
of $\approx$120 AU and $\approx$4.5~AU.
Here we present a substellar companion to the active M3.5 star 2MASS~J01225093--2439505
(hereinafter 2MASS~0122--2439).
The young age of 2MASS~0122--2439 was first recognized by \citet{Riaz:2006p20030} based on its
high X-ray luminosity and strong H$\alpha$ emission (9.7~\AA).
In their kinematic analysis of active M dwarfs, \citet{Malo:2013p24348} identified 2MASS~0122--2439 
as a likely member of the $\sim$120~Myr AB~Dor young moving group (YMG; \citealt{Zuckerman:2004p22744}).
Neither a parallax nor a radial velocity has been published for this otherwise anonymous star.
Below we describe the discovery and spectroscopic analysis of a dusty L dwarf companion 
to 2MASS~0122--2439 with an estimated mass near the deuterium-burning limit.

\section{Observations}{\label{sec:obs}}

\subsection{Subaru/IRCS Adaptive Optics Imaging}{\label{sec:subobs}}

We first imaged 2MASS~0122--2439 with the Infrared Camera and Spectrograph 
(IRCS; \citealt{Kobayashi:2000p20330}; \citealt{Tokunaga:1998p20329}) 
combined with the AO188 adaptive optics system (\citealt{Hayano:2010p25015})
at the Subaru Telescope on 2012 Oct 12 UT.
Natural guide star (NGS) adaptive optics was used with
2MASS~0122--2439~A ($R$=13.2~mag) acting as the wavefront reference.
The observations were made with the smallest plate scale 
(20~mas/pix), resulting in a field of
view of 21$''$ across the 1024~$\times$~1024~array.
We obtained short (5 exposures $\times$ 5 coadds $\times$ 1.0~s) 
and long (2 exposures $\times$ 1 coadd $\times$ 60~s)
dithered frames with the $H$ and $K$ filters (Mauna Kea Observatory system; 
\citealt{Simons:2002p20490}; \citealt{Tokunaga:2002p20495}).  
Seeing was poor (1--2$''$) and variable
throughout the night, resulting in rapid changes in the quality of AO correction.
Dome flats obtained at the start of the night were used to correct for
pixel sensitivity variations.
A faint ($\Delta$mag $\sim$ 5) point source  was resolved $\sim$1$\farcs$5
from 2MASS~0122--2439~A, but the AO correction was only only good enough
to extract precise astrometry ($<$20~mas) for a single short-exposure frame.

Astrometry for this image was measured by fitting an analytic model of three elliptical Gaussians to each 
binary component as described in \citet{Liu:2008p14548}.
As in previous work (\citealt{Dupuy:2009p18533}; \citealt{Dupuy:2010p21117}), 
we determined the errors in separation and position angle (PA) by
fitting simulated data created from the image itself.  
Images of the primary are successively scaled to the measured flux of the companion
and injected at the best-fit separation (70.75~pix),
adding random offsets with an rms of 1.0~pixel.  
We also allowed for random offsets in the flux ratios of
our injected companions (rms = 0.3\,mag), distributing them 
uniformly in PA but avoiding the $\pm10$~deg PA range where the
actual companion is located.  
The same fitting routine is applied to $10^2$ such simulated companion images.

North alignment and pixel scale was measured using dithered observations of the
young 2$\farcs$4 M/L binary 1RXS~J235133.3+312720~AB (\citealt{Bowler:2012p23851})
acquired on the same night (10 exposures $\times$ 5 coadds $\times$ 0.5 s)
with an identical setup.  
We measure a plate scale of 20.41~$\pm$~0.05 mas~pix$^{-1}$ and a North orientation
of +89$\fdg$03~$\pm$~0$\fdg$09 based on astrometry of the system
from Keck/NIRC2 from Bowler et al., in agreement 
with the IRCS plate scale of 20.53~mas pix$^{-1}$ measured by \citet{Currie:2011p21928}.
Applying our calibration to the 2MASS~0122--2439~AB system gives a separation and PA
of 1444~$\pm$~7 mas and 216.7~$\pm$~0.2 deg.
We did not correct the data for optical distortions since the relative distortion 
is small compared to our measurement errors for separations of only a few arcseconds.

\subsection{Keck~II/NIRC2 Adaptive Optics Imaging}{\label{sec:nirc2obs}}

We obtained follow-up imaging of 2MASS~0122--2439~AB using Keck~II/NIRC2 
with NGS-AO on the nights of 18 and 19 January 2013 UT, and again on 30 June 2013 UT.
The narrow camera mode was used, resulting in a field of view  of 10$\farcs$2$\times$10$\farcs$2.
Standard dark subtraction, bad pixel correction, and flat fielding was performed
as in previous work (e.g., \citealt{Bowler:2012p23851}).
For our astrometry, 
we adopt the plate scale of 9.952~mas pix$^{-1}$ and angle of +0$\fdg$252 
between the detector columns and celestial north
measured by \citet{Yelda:2010p21662}.
Details of the observations  are listed in Table~\ref{tab:astrometry}. 

\begin{figure}
  \hskip .3in
  \resizebox{7in}{!}{\includegraphics{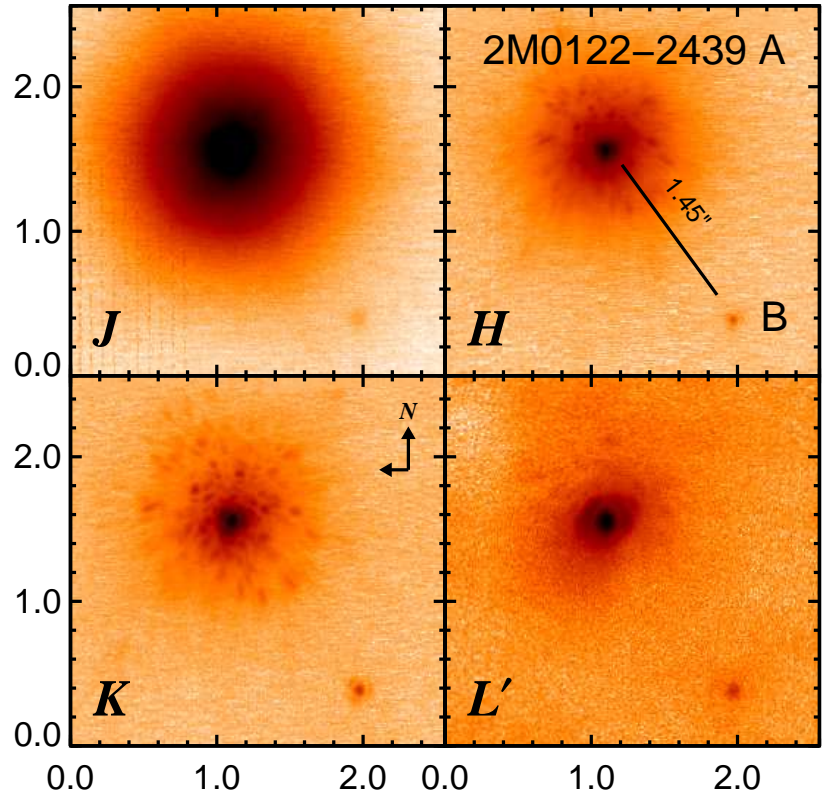}}
  \vskip -3.3in
  \caption{Keck/NIRC2 adaptive optics images of 2MASS~0122--2439~AB.  The companion  is located at a 
  projected separation of 1.45$''$ with a flux ratio of $\approx$6.2~mag in $H$.  The $J$-band data are from June 2013, 
  and the $H$, $K$, and $L'$ images are from January 2013.  North is up and East is to the left. \label{fig:imgs}  
}
\end{figure}

2MASS~0122--2439~B was detected in our $H$-, $K$-, and $L'$-band data from January 2013 and
our $J$- and $K$-band data from June 2013 (Figure~\ref{fig:imgs}), but 
not in our $J$-band images from January 2013.  For the detections, astrometry and relative
photometry was extracted either by fitting an analytic model of three elliptical Gaussians
to each component or performing aperture photometry with sky subtraction.  
The standard deviation of multiple measurements is computed to derive the PA and separation uncertainty for each filter.
A lower limit for the January 2013 $J$-band
flux ratio was computed by adding a small image of the primary star scaled to 3 times the background
rms at the known position of the companion, and then extracting 
flux ratios using the same fitting method.
The results are listed in Table~\ref{tab:astrometry}.
Although the $K$-band photometry from NIRC2 and IRCS  
disagree at the 2.2-$\sigma$ level, we note that the companion is well-resolved from
the primary in our NIRC2 data, but it sits in a large halo in the IRCS 
data because of poor seeing that night.  The flux ratio
from NIRC2 is therefore more reliable than the IRCS measurement, as indicated by
their relative uncertainties (0.04~mag vs 0.24~mag).  Moreover, our 
two epochs of $K$-band data from NIRC2 are mutually consistent.

\subsection{Keck~I/OSIRIS Near-Infrared Spectroscopy of 2MASS~0122--2439~B}{\label{sec:osirisobs}}

We targeted 2MASS~0122--2439 B with the 
OH-Suppressing Infrared Imaging 
Spectrograph (OSIRIS; \citealt{Larkin:2006p5570}) on 2013 February 4 UT
with the Keck I Telescope using NGS-AO.  
The $Hbb$ and $Kbb$ filters were used with the 50~mas plate scale,
resulting in a lenslet geometry of 16~$\times$~64, a field of view of 0$\farcs$56~$\times$ 2$\farcs$24,
and a resolving power $R$~$\equiv$~$\lambda$/$\Delta \lambda$~$\sim$~3800 (Table~\ref{tab:spectroscopy}).
The long axis of the detector was oriented perpendicular to the binary position angle
and nodded along the detector by $\sim$1$''$ in an ABBA pattern.
We obtained a total on-source integration of 35~min in $K$ and 30~min in $H$ in $\sim$1$''$
seeing conditions over an airmass of 1.7--2.3.
Immediately after our science observations, we acquired spectra of the A0V star HD~20878 at a similar airmass for
telluric correction. 

The 2D data were reduced and rectified into wavelength-calibrated 3D cubes with the OSIRIS Data Reduction
Pipeline.  A new grating was installed in OSIRIS in December 2012 so we
made use of a preliminary set of rectification matrices obtained shortly after our observing run on
2013 February 16 (J. Lyke 2013, private communication).
The spectra were extracted using aperture photometry with an aperture radius of 3~spaxels,
and were then median-combined after scaling each spectrum to a common level.
Measurement uncertainties were derived by computing the standard deviation of the scaled flux level for 
each spectral channel.
Telluric correction was performed using a generalized version of the Spextool 
reduction package for IRTF/SpeX (\citealt{Vacca:2003p497}; \citealt{Cushing:2004p501}).
Each band was then flux-calibrated using our photometry measurements in Table~\ref{tab:properties}.

\subsection{Keck~I/HIRES Spectroscopy of 2MASS~0122--2439~A}{\label{sec:hiresobs}}

We obtained an optical spectrum of 2MASS~0122--2439~A 
with the High REsolution Echelle Spectrometer (HIRES; \citealt{Vogt:1994p18398}) on the Keck I Telescope on 
2011~Dec~28~UT (HJD 2456289.72878786). 
The HIRESr setting was used with the GG475 filter and a slit width of 0$\farcs$861, producing a resolving power of $R$~=~48000
from 6300--7800 \AA \ and 7850--9200~\AA \ for the green and red chips.
The integration time was 90~s, but the S/N was somewhat low ($\sim$20 per resolution element)
as a result of poor dome seeing ($\sim$2$''$).
Wavelength calibration was achieved using Th/Ar lamps. 
Details about the data reduction and spectral extraction can be found in \citet{Shkolnik:2009p19565} 
and \citet{Shkolnik:2012p24056}.

We measured a radial velocity (RV) for 2MASS~0122--2439~A by cross-correlating the spectra from the red and green
chips with 
the M3.5 RV standard GJ~273 (\citealt{Nidever:2002p25025}) obtained on the same night.
We found an RV of 9.5~$\pm$~1.1~km s$^{-1}$ and 9.7~$\pm$~0.9~km s$^{-1}$ for the red and green chips, respectively.
The uncertainties for both chips are the quadrature sum of the following error terms: 
the rms from 6 (11) spectral orders for 2MASS~0122--2439~A for the red (green) chip (0.9~km s$^{-1}$ and 0.6~km s$^{-1}$), 
the rms from the standard GJ~273 (0.4~km s$^{-1}$), and a estimated systematic drift of $\sim$0.5~km~s$^{-1}$ during the night.
We adopt the mean value of both chips, 9.6~$\pm$~0.7~km s$^{-1}$.

\subsection{IRTF/SpeX Near-Infrared Spectroscopy}{\label{sec:irtfobs}}

As a comparison template for our spectrum of 2MASS~0122--2439 B, we present observations of the 
companion to the young M4.5 star LP~261--75 obtained with IRTF/SpeX in prism mode 
(\citealt{Rayner:2003p2588}) on 2006 November 19~UT.
LP~261--75~B (also known as 2MASS~J09510549+3558021 and NLTT~22741~B) is an L6.5 brown dwarf first
identified by \citet{Kirkpatrick:2000p14855} and found to be comoving with the 
young ($\sim$100--200~Myr) M dwarf LP~261--75~A 
by 
\citet{Reid:2006p22856}.
A total on-source integration time of 16~min was obtained in an ABBA pattern 
with the 0$\farcs$5 slit, yielding a resolving power of $\sim$150.
Arc lamps for wavelength calibration were acquired at the same telescope position.
The A0V star HD 89239 was observed at a similar airmass as the science observations
for telluric correction.
The Spextool reduction package was used to reduce, extract, and telluric-correct the observations
(\citealt{Vacca:2003p497}; \citealt{Cushing:2004p501}).

\section{Results}{\label{sec:results}}

\subsection{Distance}{\label{sec:distance}}

Since no parallax is available for 2MASS~0122--2439~A, we estimate its distance photometrically
using empirical color-magnitude relations.
This requires knowledge of the system age since pre-main sequence isochrones
span $\sim$3 mag in $M_V$ (a factor of $\sim$4 in inferred distance) from 10~Myr to the 
zero-age main sequence ($\sim$160~Myr for a 0.4~\Msun \ star; \citealt{Baraffe:1998p160}).
As discussed in Section~\ref{sec:kinematics}, the kinematics of 2MASS~0122--2439~AB are marginally consistent with
the AB Dor YMG, which has an age similar to the Pleiades (\citealt{Luhman:2005p22437}; \citealt{Barenfeld:2013p24822}).
We therefore make use of the Pleiades isochrone to compute a photometric distance for 
our target.  Using the Pleiades membership list from \citet{Stauffer:2007p19665} and
a cluster distance of 133~pc (\citealt{Soderblom:2005p23093}), 
we fit a sixth-order polynomial to the cluster sequence in $M_V$ vs. $V$--$K_S$.
Unresolved binaries will systematically bias this fit to brighter values compared to the
single star locus.  We therefore shift the distribution 0.15~mag fainter as a compromise between
a more populous single star locus and an upper envelope of equal-flux binaries, which 
will brighten sources by as much as 0.75~mag.
The resulting fit is $M_V$ = 0.15 + $\sum_{i=0}^{5}$~$c_i$$\times$($V$--$K_S$)$^i$, where
$c_i$ = \{2.15195, --2.39272, 5.95275, --3.25361, 0.850172, --0.106731, 0.00519175\}.
The rms about the fit is 0.28~mag and it is valid between $V$--$K_S$ = 0.5--6.0~mag.
Applying the relation to 2MASS~0122--2439~A gives $M_V$ = 11.5~$\pm$ 0.3~mag
and a photometric distance of 36~$\pm$~4~pc.
We assume 2MASS~0122--2439~A is single since our NIRC2 images rule out equal-flux binaries down to $\approx$50~mas,
and there is no evidence it is a double-lines spectroscopic binary from our HIRES observations.

As a comparison, the $M_V$ vs. $V$--$J$ relations for main-sequence M dwarfs 
from \citet{Lepine:2005p19458} yield $M_V$ = 11.9~$\pm$~0.2~mag and a main sequence distance estimate of 
29$^{+11}_{-8}$~pc.  Similarly, at the other plausible age extreme ($\approx$10~Myr), 
the $M_V$ vs. ($V$--$K$) polynomial fit for the $\approx$12~Myr $\beta$~Pic YMG from \citet{Riedel:2011p22580} 
implies a distance of 58~$\pm$~2~pc. (No rms is given for their fit, so the error only incorporates 
photometric uncertainties.)
We adopt the photometric distance of 36~$\pm$~4~pc from Pleiades isochrones for this work, 
which agrees with the
statistical distance estimate of 33~$\pm$~1~pc from \citet{Malo:2013p24348}; however, we note  
that distances of $\approx$30--60~pc are possible if 2MASS~0122--2439~AB is not a member of the AB~Dor group.

\subsection{Common Proper Motion}

\begin{figure}
  \vskip 1.in
  \hskip .1in
  \resizebox{7in}{!}{\includegraphics{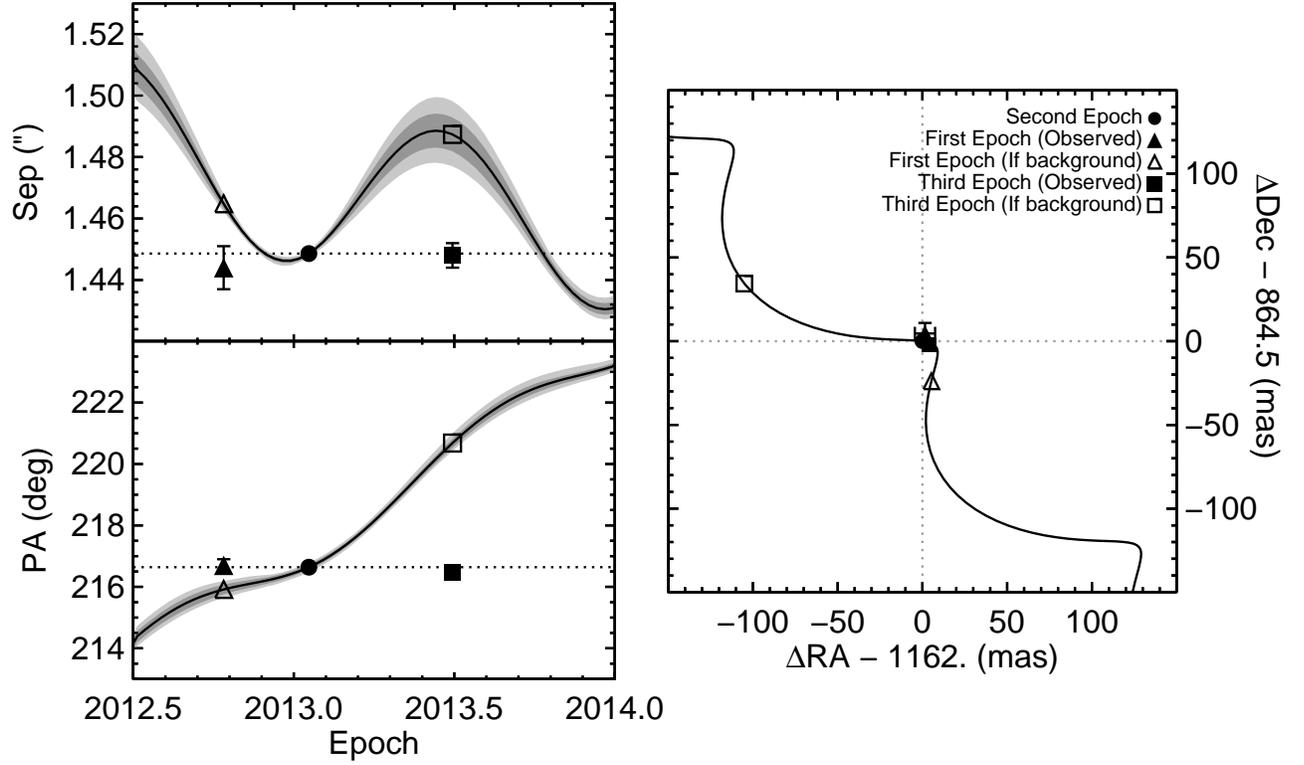}}
  \caption{Astrometric test of common proper motion for 2MASS~0122--2439~AB.  
  Black curves show the expected behavior of a 
  stationary background object based on our January 2013 NIRC2 measurement 
  for 2MASS~0122--2439~B (filled circle), the primary's proper motion, and the photometric distance (Table~\ref{tab:properties}).  
  Together our October 2012 IRCS astrometry (filled triangle) and June 2013 NIRC2 astrometry differ 
  from the stationary track in separation (upper left) and position angle (lower left) 
  by 28-$\sigma$.
  The right panel shows the relative change in right ascension and declination among
  the three epochs.
  Here $\Delta$ refers to primary minus secondary position.   \label{fig:backtracks} }
\end{figure}

The relatively large proper motion of
2MASS~0122--2439~A (170~$\pm$~2.5~mas~yr$^{-1}$ ) allows us to test whether the candidate companion is comoving 
based on our IRCS and NIRC2 AO imaging, which
span $\approx$8.5 months in time (Table~\ref{tab:astrometry}).
Figure~\ref{fig:backtracks} shows the relative proper and parallactic 
motion of a stationary background object, adopting a photometric distance
of 36~$\pm$~4~pc and the proper motion from UCAC4 listed in Table~\ref{tab:properties}.
Our NIRC2 astrometry is more precise, so we use our second epoch (Jan 2013) of data as the baseline for the
background tracks, while our first epoch (Oct 2012) IRCS and third epoch (June 2013) NIRC2 measurements are used 
to test the background hypothesis.
The predicted separation and position angle for a stationary object at epoch 2012.780 is 
1$\farcs$4648~$\pm$~0$\farcs$0011 and 215.92~$\pm$~0.13$^{\circ}$.
Our measured separation at that epoch (1$\farcs$444~$\pm$~0$\farcs$007) is 2.9-$\sigma$
from the background model and the measured PA 
(216.7~$\pm$~0.2$^{\circ}$) differs by 3.3-$\sigma$.
Likewise, for epoch 2013.495 the predicted background astrometry is 1$\farcs$487~$\pm$0$\farcs$005 and 
220.68~$\pm$~0.14$^{\circ}$, which is inconsistent with our separation and position angle measurements 
by 6.1-$\sigma$ and 27-$\sigma$, respectively.
Altogether, our astrometry rule out a background model at the 28-$\sigma$ level.
Note that if we assume a farther distance of 60~$\pm$~5~pc, which corresponds to a younger
primary age of $\approx$10~Myr, we can reject the background hypothesis at the 32-$\sigma$ level.

Bayesian techniques 
provides a straightforward way to quantitatively 
compare the relative merit of the stationary and comoving models 
(\citealt{Schwarz:1978p17783}, \citealt{Liddle:2009p19706}).   
For two competing models, the posterior odds 
$P$($M_1$$\vert$$d$)/$P$($M_2$$\vert$$d$) for
models $M_1$ and $M_2$ given the data $d$ are equal to the
Bayes factor $P$($d$$\vert$$M_1$)/$P$($d$$\vert$$M_2$)
times the prior odds of each model $P$($M_1$)/$P$($M_2$).
For simplicity, we assume equal prior odds for both models.
Assuming normally-distributed data, the Bayes factor simplifies to the 
likelihood ratio $e^{-\Delta \chi^2 /2}$.  
Here  
\begin{displaymath}
\chi^2 = \sum_{i=1}^{N-1} \left( \frac{(\theta_{meas, i} - \theta_{pred, i})^2}{\sigma_{\theta, meas, i}^2 + \sigma_{\theta, pred, i}^2} + 
 \frac{(\rho_{meas, i} - \rho_{pred, i})^2}{\sigma_{\rho, meas, i}^2 + \sigma_{\rho, pred, i}^2} \right),
\end{displaymath}
where $\theta$, $\rho$, and $\sigma$ are the measured and predicted 
PA, separation, and their associated uncertainties at epoch $i$ for $N$ epochs of astrometry.
The reduced $\chi^2$ values (2 degrees of freedom) are 1.5 and 390 for the comoving and background scenarios. 
The Bayes factors for the comoving vs.~stationary  
models correspond to posterior odds of log($P$($M_1$)/$P$($M_2$)) = 168, indicating a decisive preference for common proper motion.

\subsection{Age and Kinematics}{\label{sec:kinematics}}

Multiple lines of evidence point to a young age for the 2MASS~0122--2439~AB system.
The primary was detected in the  $ROSAT$ (\citealt{Voges:1999p22945})
and the \emph{Galaxy Evolution Explorer} ($GALEX$; \citealt{Martin:2005p23310}; \citealt{Morrissey:2007p22251}) 
all-sky X-ray and UV surveys, indicating it has an active corona and chromosphere.
High-energy emission is associated with youth for early-M dwarfs (e.g., \citealt{Shkolnik:2009p19565}; 
\citealt{Findeisen:2010p20370}; \citealt{Irwin:2011p22865}),
though it can also be generated from close, tidally-interacting binaries 
(e.g., \citealt{Shkolnik:2010p20931}).  However, our HIRES data show no evidence that 2MASS~0122--2439~A is an SB2, 
so the activity is probably an age-related phenomenon.
We compute an X-ray luminosity of log~$L_X$=28.7~$\pm$~0.2 erg s$^{-1}$ and
a relative X-ray luminosity of log~($L_X$/$L_{\odot}$)~=~--3.17~$\pm$~0.3~dex
using the measured $ROSAT$ count rate (0.0570~$\pm$~0.0142 cnt s$^{-1}$), the 
count rate-to-flux conversion from \citet{Fleming:1995p23307}, and our distance estimate.
The X-ray luminosity of 2MASS~0122--2439~A is comparable to known YMG members and more active
field objects (\citealt{Preibisch:2005p330}; \citealt{Bowler:2012p23980}).
The X-ray luminosity is near the saturation level of $\sim$--3~dex (\citealt{Riaz:2006p20030}).

Young M dwarfs produce a wealth of chromospheric emission lines, which decrease with age
as magnetic dynamo strengths fade 
(e.g., \citealt{Delfosse:1998p23131}; \citealt{Wright:2011p23132}).  As a result, $GALEX$ has become
an excellent resource to identify nearby low-mass members of YMGs
(\citealt{Shkolnik:2011p21923}; \citealt{Rodriguez:2011p21813}; 
\citealt{Shkolnik:2012p24056}; \citealt{Schlieder:2012p23477}).
The $GALEX$ $NUV$--$J$ and $FUV$--$J$ colors of 2MASS~0122--2439~A are 10.59~$\pm$~0.14~mag
and 11.1~$\pm$~0.3~mag, respectively.  Compared to YMG 
members of various ages from \citet{Findeisen:2011p22756}, 
2MASS~0122--2439~A 
lies redward of the Hyades sequence ($\sim$600~Myr) in a region occupied by members of the
TW~Hya ($\sim$8~Myr), $\beta$~Pic~($\sim$12~Myr), Tuc-Hor ($\sim$30~Myr), and AB~Dor ($\sim$120~Myr)
associations.
Because of the large scatter in these relations, 
 the high energy activity provides only a coarse age constraint of $\lesssim$500~Myr.
We measure an H$\alpha$ $EW$ of 
--5.8~$\pm$~0.5~\AA \ for the primary 
from our HIRES spectrum, which
points to an age less than $\sim$3~Gyr based on
M dwarf activity lifetimes (\citealt{West:2008p19562}).
Note that \citet{Riaz:2006p20030} find a value of 
--9.7~\AA, indicating some variability is present\footnote{We note that the 
\emph{Wide-field Infrared Survey Explorer} ($WISE$; \citealt{Wright:2010p22018})
$W1$--$W4$ color of 2MASS~0122--2439~A (0.8~$\pm$~0.2~mag)
may suggest a weak 22~$\mu$m excess compared to early-M dwarfs
in the field, which have colors of $\approx$0.3~mag (\citealt{Avenhaus:2012p25021}).
However, the significance of the 22~$\mu$m detection is only 2.5~$\sigma$ so follow-up observations
are needed for verification.}.

\begin{figure}
  \vskip 1in
  \hskip .2in
  \resizebox{6.5in}{!}{\includegraphics{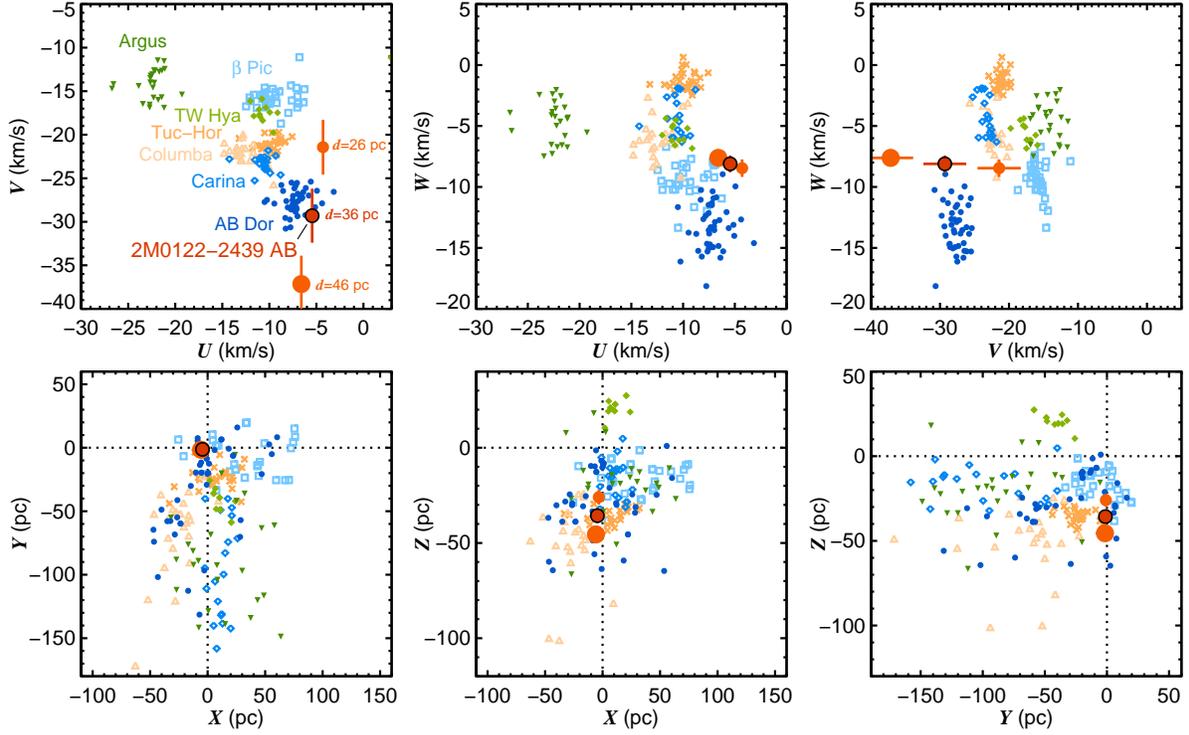}}
  \vskip -0.2in
  \caption{$UVW$ kinematics and $XYZ$ space positions for 2MASS~0122--2439~AB relative to young moving groups in the
  solar neighborhood from \citet{Torres:2008p20087}.  
  2MASS~0122--2439~AB is consistent with the AB Dor moving group ($\approx$120~Myr) 
  in $U$ and $V$ but only
  marginally consistent with known members in $W$.  We tentatively associate 2MASS~0122--2439~AB with
  AB Dor, but a parallax is needed for verification. The red circle shows our photometric distance of 36~$\pm$~4~pc,
  and the two orange circles show distances of 26 and 46 pc for comparison.
  Errors in kinematics and space positions for 2MASS~0122--2439~AB incorporate
  uncertainties in the proper motion, radial velocity, and photometric distance.
  \label{fig:uvw} } 
\end{figure}

\citet{Malo:2013p24348} identify 2MASS~0122--2439 as a likely member of the 
AB~Dor YMG (\citealt{Zuckerman:2004p22744}) 
based on its proper motion, photometry, and 
sky position.   
They predict 
an RV of 15.5~$\pm$~2.1~km s$^{-1}$, 
but our measured value of 9.6~$\pm$~0.7~km s$^{-1}$ differs from this by 2.8~$\sigma$.
Using the proper motion, RV, and photometric distance of  2MASS~0122--2439~A, we 
compute $UVW$ space velocities and $XYZ$ galactic positions (Table~\ref{tab:properties}),
which are 
shown in Figure~\ref{fig:uvw} relative
to YMG members from \citet{Torres:2008p20087}.
2MASS~0122--2439~AB agrees with the AB~Dor group in $U$ and $V$,
but is near the outskirts of known members in $W$.
We therefore tentatively adopt the age of the AB~Dor YMG for this work
($\sim$120~Myr; \citealt{Luhman:2005p22437}; \citealt{Barenfeld:2013p24822}), but
note that other ages are possible if 
a future parallax measurement shows it does not belong to that moving group.

The lack of  \ion{Li}{1} $\lambda$6708 absorption in 2MASS~0122--2439~A ($EW$$<$50~m\AA)
from our HIRES spectrum
provides a strict lower limit on the age of the system since the Li depletion boundary 
is a strong function of temperature and age.
Compared to 
the Li equivalent widths for cool YMG members from  \citet{Mentuch:2008p23922},
2MASS~0122--2439~A must be older than the TW~Hydrae moving group ($\sim$8~Myr) 
based on its spectral type of M3.5.
Lithium 
depletion predictions from evolutionary models can also be used 
to set a lower age limit on the system.
For 2MASS~0122--2439~A, the models of \citet{Chabrier:1996p23529} indicate a Li-burning
timescale of $\sim$20--30~Myr.  The accretion history 
of young stars can result in more rapid depletion times (\citealt{Baraffe:2010p21792}),
so 20--30~Myr is effectively the oldest possible lower limit for the system age.
However, the spectral shape of 2MASS~0122--2439~B implies an approximate upper age limit
of $\sim$120~Myr (see Section~\ref{sec:spec}).

\subsection{Luminosity, Mass, and Effective Temperature}{\label{sec:lum}}

For the primary 2MASS~0122--2439~A, we derive a luminosity of
log~($L/L_{\odot}$) = --1.72~$\pm$~0.11~dex using the $H$-band
bolometric correction from \citet{Casagrande:2008p23483} and the 
photometric distance from Section~\ref{sec:distance}.
At an age of 120~Myr (10~Myr), the luminosity of 2MASS~0122--2439~A
implies a mass of 0.40~$\pm$0.05~$M_{\odot}$
(0.13~$\pm$0.02~$M_{\odot}$) and an effective temperature of 
3530~$\pm$~50~K (3150~$\pm$~40~K)
based on the solar-metallicity evolutionary models of \citet{Baraffe:1998p160}.

We calculate the luminosity of 2MASS~0122--2439~B by
integrating our flux-calibrated
1.5--2.4~$\mu$m spectrum (Section~\ref{sec:spec}) together with a scaled model spectrum
at shorter (0.001--1.5~$\mu$m) and longer (2.4--1000~$\mu$m) wavelengths.
Based on the mid-L spectral type of 2MASS~0122--2439~B (Section~\ref{sec:spec}), 
we use a solar metallicity BT-Settl synthetic spectrum 
(\citealt{Allard:2010p21477})
with an effective
temperature of 1700~K and a log-gravity of 4.5~dex (cgs), from which we find a luminosity
of log~$L/L_{\odot}$ = --4.19~$\pm$~0.10~dex
using our photometric distance.
Spectral and photometric (flux calibration) 
measurement uncertainties together with our photometric distance estimate 
are taken into account in a Monte Carlo fashion. 
Note that even though our spectroscopic data only contribute  32\% of the 
bolometric luminosity, 
the choice of the model spectrum for the bolometric correction 
has only a minor influence on the total luminosity since the models are
scaled to the flux-calibrated spectra.
For example, the \{1500~K, 4.5~dex\} spectrum yields a consistent 
luminosity of --4.16~$\pm$0.10~dex.

Figure~\ref{fig:hrd} shows the luminosity and age of 2MASS~0122--2439~B compared to the 
cloudless (dotted gray tracks) and cloudy (solid gray tracks) ``hot start'' cooling models 
of \citet{Saumon:2008p14070}.  2MASS~0122--2439~B sits in a region 
at the deuterium-burning limit where mass tracks overlap because the onset
of deuterium burning is a function of mass   
(e.g., \citealt{Burrows:2001p64}; \citealt{Spiegel:2011p22104}).
Its position coincides with the 0.012--0.013~\Msun \ ($\sim$12--14~\Mjup) and 
0.022--0.026 \Msun \ ($\sim$23--27~\Mjup) cloudy tracks.
These mass ranges appear to be more likely for 2MASS~0122--2439~B than the
15--20~\Mjup \ range, which lies at slightly lower luminosities and younger ages.
This highlights an important (but often unrecognized) feature 
for \emph{all} hot-start evolutionary models: in this strip near the deuterium-burning limit, 
a luminosity and age alone do not coincide with a unique mass.  
However, as we discuss Section~\ref{sec:disc}, a detailed spectroscopic comparison of objects
at the same location might break this mass degeneracy.
Here we have assumed the system belongs to the AB~Dor YMG, but if it is not a member then the age
may be as low as $\approx$10~Myr, in which case the corresponding luminosity and mass 
of 2MASS~0122--2439~B are --3.66~$\pm$~0.08~dex and $\approx$11~\Mjup.

The formation and accretion history of substellar companions can strongly influence the evolution
of their luminosity up to ages of $\sim$1~Gyr (\citealt{Marley:2007p18269}; \citealt{Fortney:2008p8729}; 
\citealt{Spiegel:2012p23707}).  
Our hot-start mass estimates, which are based on arbitrarily high
initial specific entropies, are in fact lower limits for the actual mass of 2MASS~0122--2439~B 
if some of the initial gravitational energy is radiated away as accretion luminosity (\citealt{Marleau:2013p24445}).
In Figure~\ref{fig:coldhrd} we consider the ``cold start'' evolutionary models of \citet{Molliere:2012p25183}, which
are based on a core accretion formation scenario.  
Although core accretion is unlikely to have created such a massive companion at the present location
of 2MASS~0122--2439~B (52~AU in projected separation), it remains possible that dynamical interactions 
with another close-in companion 
could have scattered it to a wide orbit (e.g., \citealt{Veras:2009p19654}).
Above $\approx$13~\Mjup, deuterium burning results in a luminosity ``bump'' which onsets
at progressively later ages for lower masses.  
For 2MASS~0122--2439~B, deuterium burning in the cold start models results in dual-valued masses 
of $\approx$14~\Mjup \  and $\approx$23~\Mjup, 
which are similar to those from the hot start models.

Hot start evolutionary models predict effective temperatures between $\sim$1350--1500~K
based on the luminosity and age of 2MASS~0122--2439~B (Figure \ref{fig:twom0122_hrd_teff} ), 
differing somewhat for the two possible mass regimes.
At lower masses of $\sim$13~\Mjup, temperatures from the \citet{Saumon:2008p14070} models range from
1350--1410~K for the cloudless, cloudy, and hybrid prescriptions.  At higher masses of $\sim$24~\Mjup, 
the temperatures are between 1450---1500~K.

\subsection{Spectral Properties of 2MASS~0122--2439 B}

\subsubsection{Photometry}

$J$-, $H$-, and $K$-band photometry for 2MASS~0122--2439~B is computed using our measured
relative photometry and the apparent magnitudes of 2MASS~0122--2439~A from 2MASS (\citealt{Skrutskie:2006p589}),
converted to the MKO filter system based on the 
relations from \citet{Leggett:2006p2674}.
The $L'$ photometry of 2MASS~0122--2439~B is derived assuming a $K$--$L'$ color of 0.2~$\pm$~0.1~mag
for the primary, which is the typical value for an M3.5 dwarf (\citealt{Golimowski:2004p15703}).
The results are listed in Table \ref{tab:properties}.

\begin{figure}
  \hskip -1.2in
  \resizebox{10in}{!}{\includegraphics{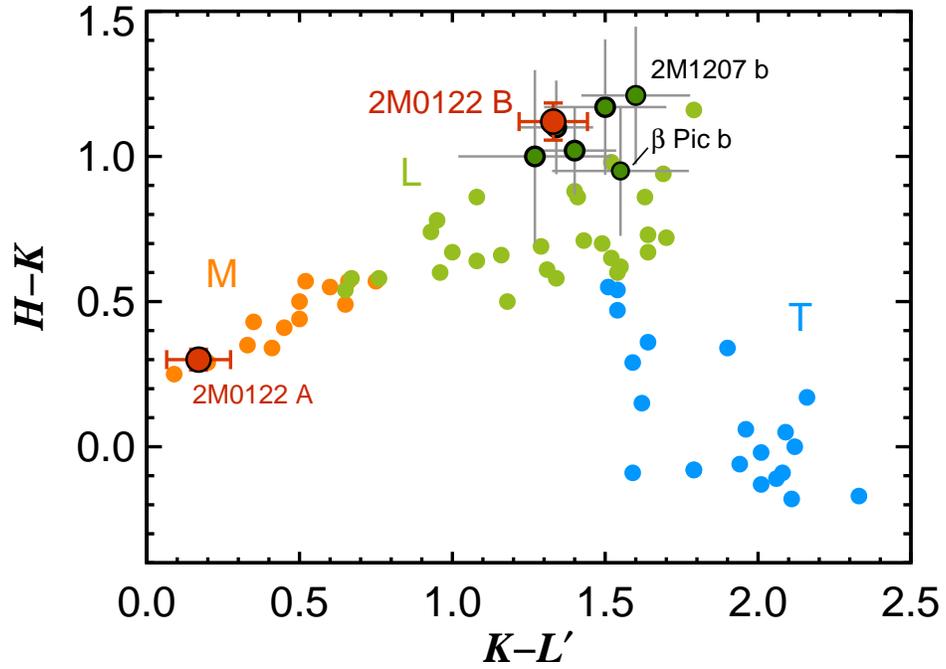}}
  \vskip -1.5in
  \caption{Color-color diagram showing the position of 2MASS~0122--2439~A and B relative to
  M (orange), L, (light green) and T (blue) dwarfs in the field.
  2MASS~0122--2439~B is particularly red in $H$--$K$ compared to field objects, better resembling 
   the planetary-mass companions HR~8799~bcde, $\beta$~Pic~b, and  2MASS~1207--3932~b (green circles). 
   Photometry for field objects is from \citet{Leggett:2010p20094}, and photometry for the planetary-mass
   companions is from \citet{Skemer:2012p24037}, \citet{Bonnefoy:2013p24399}, and \citet{Mohanty:2007p6975}.
   All photometry is in the MKO filter system.
        \label{fig:hklccd} } 
\end{figure}

Figure~\ref{fig:hklccd} shows the 
 $H$--$K$ vs $K$--$L'$ colors of 2MASS~0122--2439 B 
compared to field M, L, and T dwarfs (\citealt{Leggett:2010p20094})
and the planetary-mass companions HR~8799~bcde, 2MASS~1207--3932~b,
and $\beta$~Pic~b.
The $K$--$L'$ color of 2MASS~0122--2439 B is similar to mid-L dwarfs,
but the $H$--$K$ value is redder than most field objects by $\sim$0.3~mag.

\begin{figure}
  \vskip 1.in
  \hskip -.2in
  \resizebox{8in}{!}{\includegraphics{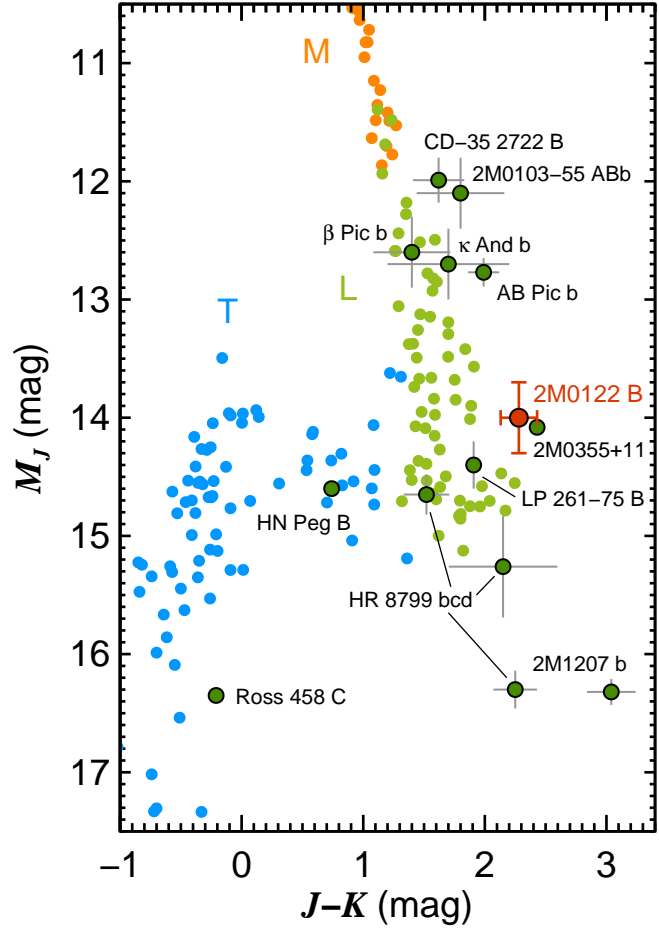}}
  \vskip -0.1in
  \caption{$M_J$ vs. $J$--$K$ diagram for field M (orange), L (light green), and T (blue) dwarfs. 
  Dark green circles show substellar companions with parallactic distances (either to the primaries or the
  secondaries themselves).  For comparison we also overplot the young, dusty L5 member of the AB Dor YMG
   2MASS~0355+1133 (\citealt{Cruz:2009p19453}; \citealt{Faherty:2012p24307}; \citealt{Liu:2013p25024}).
 The $M_J$ magnitude and $J$--$K$ color of 2MASS~0122--2439~B is similar to 2MASS~0355+1133.
     Photometry is from \citet{Dupuy:2012p23924}, \citet{Skemer:2012p24037},  \citet{Carson:2013p24396}, \citet{Delorme:2013p25184}, 
     and \citet{Bonnefoy:2013p24399}.  All photometry is in the MKO filter system.
      \label{fig:jkcmd} } 
\end{figure}

\begin{figure}
  \vskip 1.in
  \hskip -.5in
  \resizebox{8in}{!}{\includegraphics{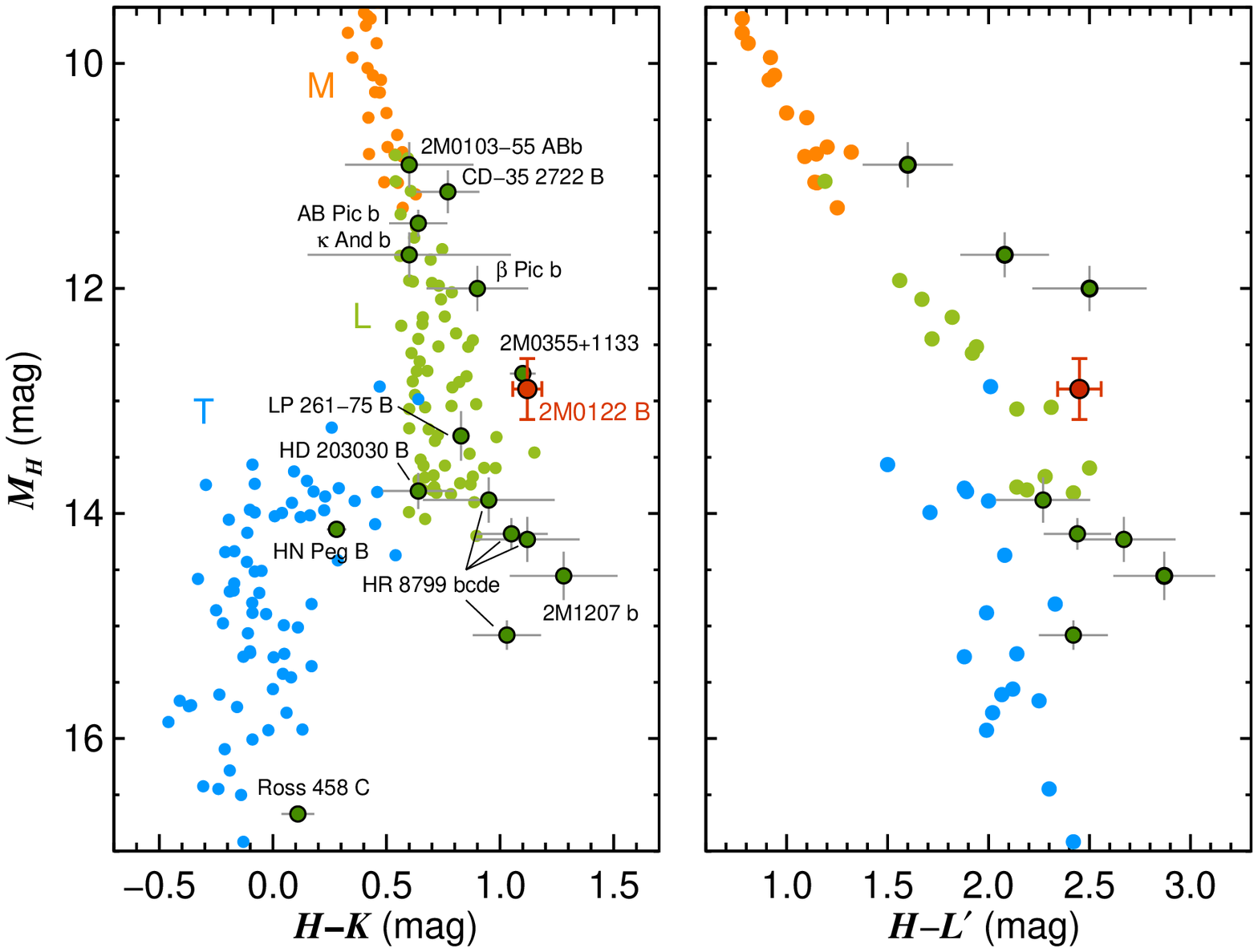}}
  \vskip -0.1in
  \caption{Similar to Figure~\ref{fig:jkcmd} but for $M_H$ vs. $H$--$K$ (left) and $M_H$ vs. $H$--$L'$ (right).
    2MASS~0122--2439~B is redder in both $H$--$K$ and $H$--$L'$ compared to the field sequence,
  indicative of a young and/or dusty atmosphere (e.g., \citealt{Liu:2013p25024}).  
     Photometry is from \citet{Metchev:2006p10342}, \citet{Dupuy:2012p23924}, \citet{Skemer:2012p24037}, \citet{Carson:2013p24396}, \citet{Delorme:2013p25184}, 
     and \citet{Bonnefoy:2013p24399}.  All photometry is in the MKO filter system.
      \label{fig:cmd} } 
\end{figure}

In Figures~\ref{fig:jkcmd} and \ref{fig:cmd} we compare the colors and absolute magnitudes
of 2MASS~0122--2439~B (based on its photometric distance) to field M, L, and T dwarfs from \citet{Dupuy:2012p23924} 
and young substellar companions with parallaxes (see the figure captions for details).  
2MASS~0122--2439~B is similar to mid- to late-L dwarfs in absolute magnitude, but is 
significantly redder than field objects and most known companions; instead its infrared colors better
resemble those of the HR~8799 planets.
Compared to young substellar companions, 2MASS~0122--2439~B has the second reddest $J$--$K$ color
after the planetary-mass companion 2MASS~J1207--3932~b (\citealt{Chauvin:2004p19400}; \citealt{Mohanty:2007p6975}).
The absolute magnitudes and colors of 
2MASS~0122--2439~B are remarkably similar to those of 
2MASS~J03552337+1133437 (hereinafter 2MASS~0355+1133), 
a young dusty L5 member of the AB~Dor moving group (\citealt{Reid:2008p20073}; 
\citealt{Cruz:2009p19453}; \citealt{Faherty:2012p24307}; \citealt{Liu:2013p25024}).
Figures~\ref{fig:jkcmd} and \ref{fig:cmd} also emphasize the uniqueness of 2MASS~0122--2439~B;
its cool temperature and very red colors make it the only companion currently known with atmospheric properties
intermediate between $\beta$~Pic~b and the HR~8799 planets.

\subsubsection{Spectroscopy}{\label{sec:spec}}

\begin{figure}
  \vskip 1.in
  \hskip -.7in
  \resizebox{8.5in}{!}{\includegraphics{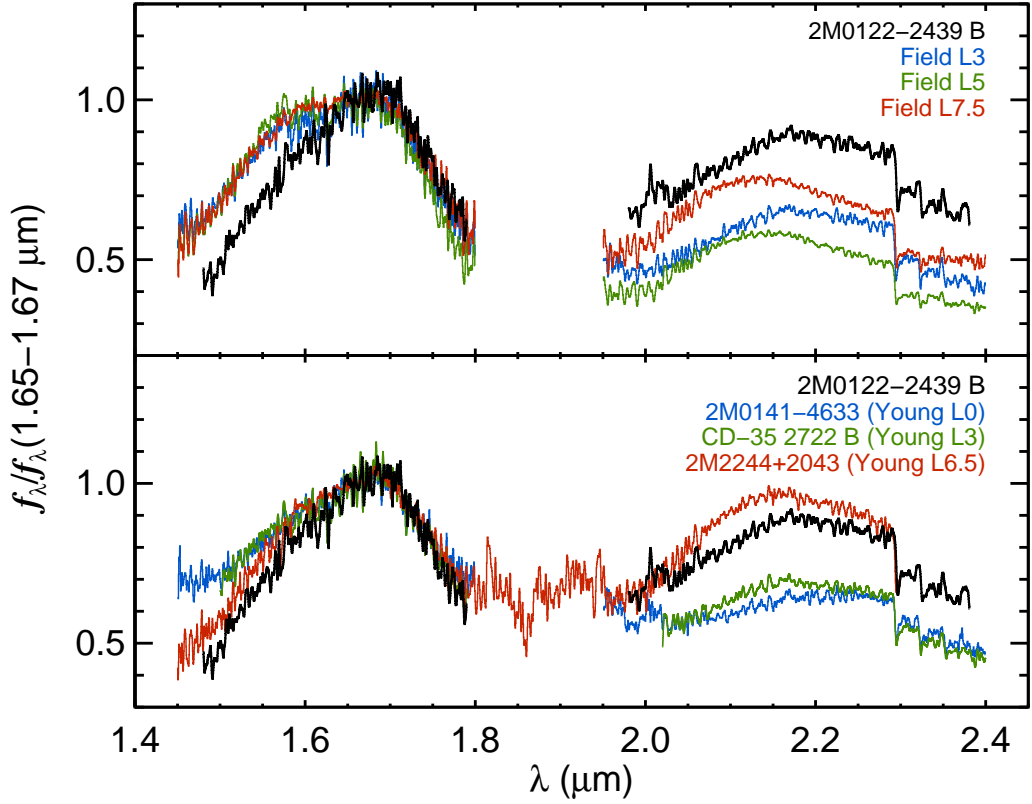}}
  \vskip -0.5in
  \caption{Comparison of 2MASS~0122--2439~B with field (top) and young (bottom) L dwarfs.
  Field objects show prominent FeH absorption between 1.57--1.62~$\mu$m and
  a wider $H$-band shape.  2MASS~0122--2439~B better resembles young
  (30--120~Myr) L dwarfs, which exhibit more angular $H$-band shapes as a result of
  gravity-sensitive steam absorption and collision-induced $H_2$ 
  absorption (\citealt{Barman:2011p22098}).  2MASS~0122--2439~B is redder in $H$--$K$ than the low-gravity
  L0 field object 2MASS~J01415823--4633574 and the L3 AB~Dor member CD--35~2722~B 
  ($\sim$120~Myr), but slightly bluer than the low-gravity L6.5 object 2M2244+2043.
  The field spectra of 2MASS~J1506+1321 (L3), SDSS~J0539--0059 (L5), and 2MASS~J0825+2115 (L7.5)   
  are from the IRTF  Spectral Library (\citealt{Cushing:2005p288}).
  The young comparison objects are from \citet{Kirkpatrick:2006p20500}, \citet{Wahhaj:2011p22103}, and
  \citet{McLean:2003p3912}.
  All spectra have been smoothed to $R$$\sim$1000 and normalized between 1.65--1.67~$\mu$m.  \label{fig:speccomp} } 
\end{figure}

\begin{figure}
  \vskip -1in
  \hskip .5in
  \resizebox{6.5in}{!}{\includegraphics{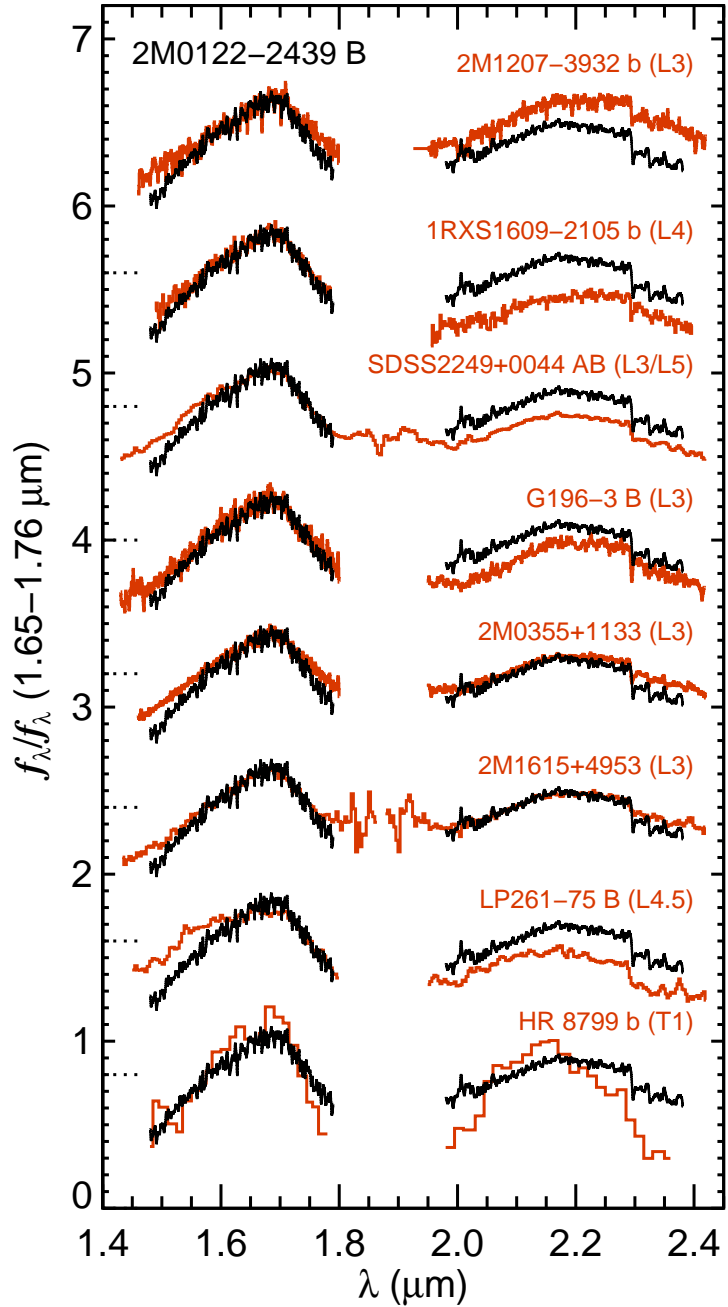}}
  \vskip -0.3in
  \caption{Our OSIRIS spectra of 2MASS~0122--2439~B compared with low-gravity L dwarfs.
  The $H$-band shape best resembles the young $\sim$L4 companion 1RXS~160929.1--210524~b, 
 the young 20--85~Myr L3 companion G196-3~B, the young field L3 dwarf 2MASS~1615+4953, 
 and the $\approx$30~Myr $\sim$T1 planet HR~8799~b.  Although it has similar colors to the 
  AB Dor member 2MASS~0355+1133,  the $H$-band shape of 2MASS~0122--2439~B appears more peaked, 
  suggesting a younger age or lower temperature for 2MASS~0122--2439~B.
  Several companions have low-gravity classifications in \citet{Allers:2013p25081} based on their spectral typing scheme:
  SDSS~2249+0044~AB is listed as intermediate gravity, G196-3~B is very low-gravity, 2MASS~0355+1133 is 
  very low-gravity, and 2MASS~1615+4953 is very low-gravity.  Note that spectral classifications listed in the figure are near-infrared spectral types.
  Spectra are from \citet{Patience:2010p21350}, \citet{Lafreniere:2008p14057}, \citet{Allers:2010p20499}, 
  \citet{Allers:2013p25081}, and \citet{Barman:2011p22098}.
     \label{fig:youngspeccomp} } 
\end{figure}

In Figure~\ref{fig:speccomp} we compare our flux-calibrated spectrum of 2MASS~0122--2439~B to
old L dwarfs in the field (top panel) and those exhibiting signs of youth (bottom panel).
The $H$-band shape of 2MASS~0122--2439~B is significantly more angular than field objects,
a prominent signature of low surface gravity (e.g., \citealt{Lucas:2001p22099}; 
\citealt{Allers:2007p66}; \citealt{Allers:2013p25081})
caused by diminished steam and collision-induced H$_2$ absorption (\citealt{Barman:2011p22098}).
2MASS~0122--2439~B appears significantly later than the young field L0 object 2MASS~J01415823--4633574 
(\citealt{Kirkpatrick:2006p20500}) and the L3 companion to the AB~Dor member CD--35~2722 
(\citealt{Wahhaj:2011p22103}), but somewhat earlier than the young L6.5 
brown dwarf 2MASSW~J2244316+204343 
(\citealt{Dahn:2002p13692}; \citealt{McLean:2003p3912}; \citealt{Kirkpatrick:2008p14118}).
Compared to other young L~dwarfs in Figure~\ref{fig:youngspeccomp},
2MASS~0122--2439~B has a similar $H$ band shape to the 5--10~Myr L4 companion 
1RXS~160929.1--210524~b (\citealt{Lafreniere:2010p20986}), 
the $\sim$20--100~Myr L3 companion G196-3~B (\citealt{Rebolo:1998p19498}; \citealt{Kirkpatrick:2008p14118}; \citealt{ZapateroOsorio:2010p20810}), 
and the young field L3 dwarf 2MASS~J1615425+495321 (\citealt{Cruz:2007p19477}; \citealt{Kirkpatrick:2008p14118}; 
\citealt{Geiler:2011p22105}; \citealt{Allers:2013p25081}).
We note that despite its similar colors to the young ($\sim$120~Myr) L5 object 2MASS~0355+1133, 2MASS~0122--2439~B
exhibits a more angular $H$-band shape and deeper CO absorption bands,
pointing to a younger age and/or lower temperature.
Unfortunately, we cannot use the new index-based spectral classification scheme of \citet{Allers:2013p25081}
since our wavelength coverage is not wide enough for their $H$-band index.
Altogether, we adopt a NIR spectral type of L4--L6 based on these relative comparisons,
although spectroscopy including the $J$ band is needed for a firmer classification.

\subsection{Properties of LP~261-75~B}

LP~261-75~B is an L6.5 (optical type) common proper motion companion 
to the active M4.5 star LP~261-75 (\citealt{Reid:2006p22856}).
Reid \& Walkowicz infer an age of 100--200~Myr based on the primary star's coronal activity.
At this age, evolutionary models predict a mass of $\approx$20~\Mjup \ for the companion,
making it a valuable benchmark system at an intermediate age to compare
with 2MASS~0122--2439~B.
Here we discuss the properties of LP~261-75~B based on our 
SpeX prism spectrum (Section~\ref{sec:irtfobs}) and a new parallactic distance 
to the system.

We calculate the luminosity of the young companion LP~261-75~B using the same method for 
2MASS~0122--2439~B in Section~\ref{sec:lum}.
Our SpeX prism spectrum is first flux-calibrated to the 2MASS $H$-band magnitude  
(15.90~$\pm$~0.14~mag). 
Using a solar metallicity \{$\Teff$=1500~K, log~$g$=4.5~dex\} BT-Settl 
spectrum for a bolometric correction together with the parallactic distance of 32.95$^{+2.80}_{-2.40}$~pc
(Vrba et al. 2014, in preparation; updated from \citealt{Vrba:2004p20827}),
we measure a bolometric luminosity of --4.43~$\pm$~0.09~dex.  
Note that our 0.8--2.45~$\mu$m spectrum represents 51\% of the bolometric flux, but the model choice for
the bolometric correction only has a modest influence on the bolometric luminosity ($\sim$0.03~dex).
As shown in Figure~\ref{fig:hrd}, the luminosity and age (100--200~Myr; \citealt{Reid:2006p22856}) 
of LP~261-75~B is consistent with 
$\sim$13~\Mjup \ and $\sim$22~\Mjup \  based on a hot start formation.  

With our new prism spectrum we also compute precise colors for LP~261-75~B
since the detections from 2MASS are near the survey limit.  Synthetic photometry based on Monte Carlo
realizations of our spectrum yields  the following colors: 
($J$--$H$)$_\mathrm{MKO}$ = 1.081~$\pm$~0.003~mag, ($H$--$K$)$_\mathrm{MKO}$ = 0.828~$\pm$~0.002~mag, 
($J$--$K$)$_\mathrm{MKO}$ = 1.908~$\pm$~0.002~mag, ($J$--$H$)$_\mathrm{2MASS}$ = 1.239~$\pm$~0.002~mag, 
($H$--$K_S$)$_\mathrm{2MASS}$ = 0.779~$\pm$~0.002~mag, ($J$--$K_S$)$_\mathrm{2MASS}$ = 2.018~$\pm$~0.002~mag.
Finally, we compute a NIR spectral type using the classification scheme from \citet{Geballe:2002p19463}.
The 1.5~$\mu$m H$_2$O and 2.2~$\mu$m CH$_4$ indices yield L3.0~$\pm$~1.0 and L6~$\pm$~1.0 classifications;
the weighted average of these is L4.5~$\pm$~0.7.
In a large analysis of young brown dwarf near-infrared spectra, \citealt{Allers:2013p25081} found that 
several index-based classification schemes are relatively insensitive to gravity.
Among these is the ``H$_2$OD'' index from \citet{McLean:2003p3912}, which is applicable for L0--L8 types.
Applying the updated index widths from Allers \& Liu yields L4.5~$\pm$~0.8,
where the uncertainty incorporates measurement errors and the rms from the relation.
The gravity score for LP~261-75~B from the Allers \& Liu scheme is 2, indicating a
high gravity (``FLD-G'') comparable to old field objects.
Altogether we adopt a near-infrared spectral type of L4.5~$\pm$~1.0 for LP~261-75~B.

\section{Discussion}{\label{sec:disc}}

\begin{figure}
  \hskip -1.5in
  \resizebox{10.5in}{!}{\includegraphics{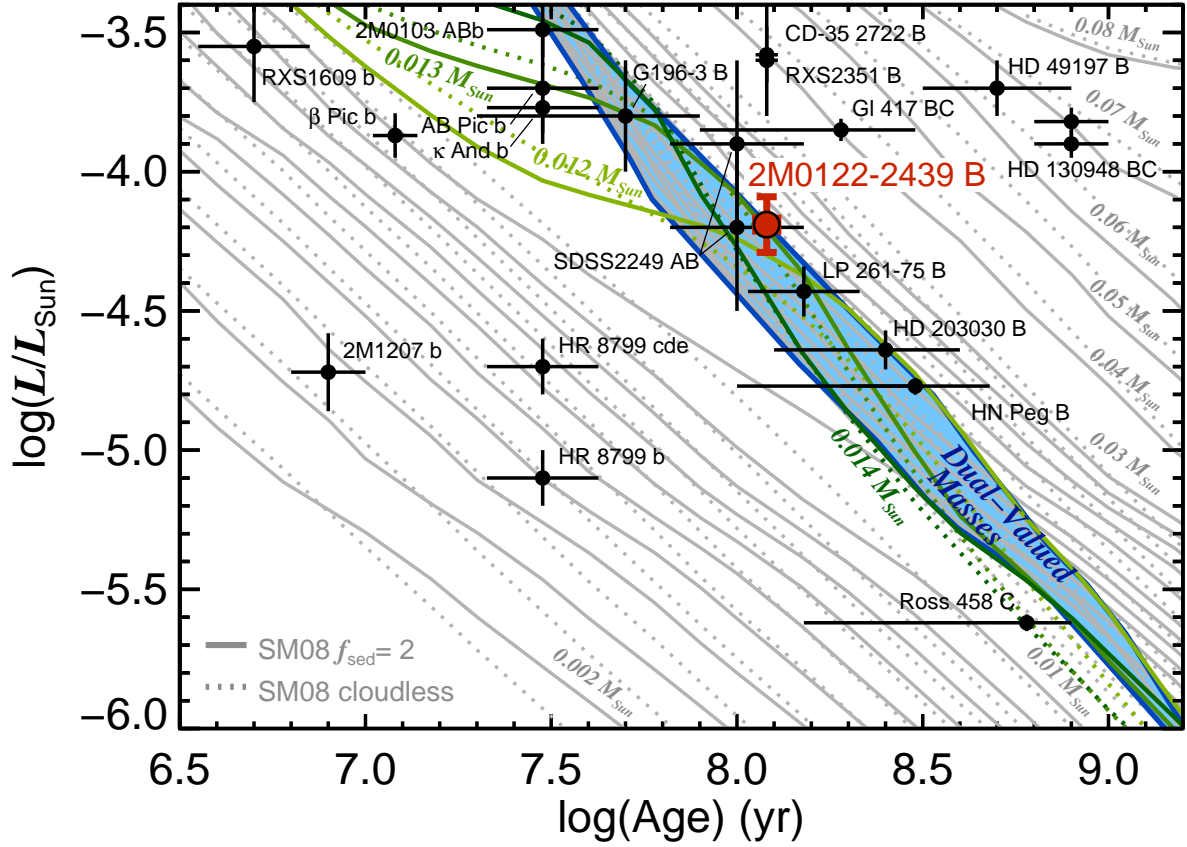}}
  \vskip -1.2in
  \caption{Luminosity and age of 2MASS~0122--2439~B compared with hot start cooling tracks for
  cloudy ($f_\mathrm{sed}$=2, solid gray curves) and clear (dotted gray curves) atmospheres 
  from the evolutionary models 
  of \citet{Saumon:2008p14070}.  2MASS~0122--2439~B and several published objects fall 
  in the region near the deuterium-burning limit
  where cooling tracks overlap as a result of mass-dependent deuterium-burning timescales (blue).
    2MASS~0122--2439~B is consistent with both $\approx$13~\Mjup \  and $\approx$25~\Mjup \ model tracks.
   Overplotted are the masses and ages of other young ($<$1~Gyr) substellar companions.
   For most objects, ages and luminosities are compiled from the literature 
   (\citealt{Lafreniere:2010p20986}; \citealt{Marois:2008p18841}; \citealt{Marois:2010p21591}; 
   \citealt{Bonnefoy:2010p20602}; \citealt{Bonnefoy:2013p24399}; \citealt{Mohanty:2007p6975}; 
   \citealt{Metchev:2006p10342}; \citealt{Luhman:2007p10341}; \citealt{Burgasser:2010p21472}; 
   \citealt{Wahhaj:2011p22103}; \citealt{Bowler:2012p23851}; \citealt{Allers:2010p20499}; \citealt{Dupuy:2010p21117})
   or measured in this work (LP~261-75~B).
   For the rest (2M0103--5515~ABb, $\kappa$~And~b, G196-3~B, HD~49197~B, Gl~417~BC), 
   luminosities are computed using distance estimates, $H$-band magnitudes,
   spectral types, and bolometric corrections using the \citet{Liu:2010p21195} relations (\citealt{Delorme:2013p25184}; 
   \citealt{Carson:2013p24396}; \citealt{ZapateroOsorio:2010p20810}; \citealt{Reid:2006p22856}; 
   \citealt{Metchev:2004p18272}; \citealt{Kirkpatrick:2001p18276}).
    \label{fig:hrd} } 
\end{figure}

\begin{figure}
  \hskip -1.5in
  \resizebox{10.5in}{!}{\includegraphics{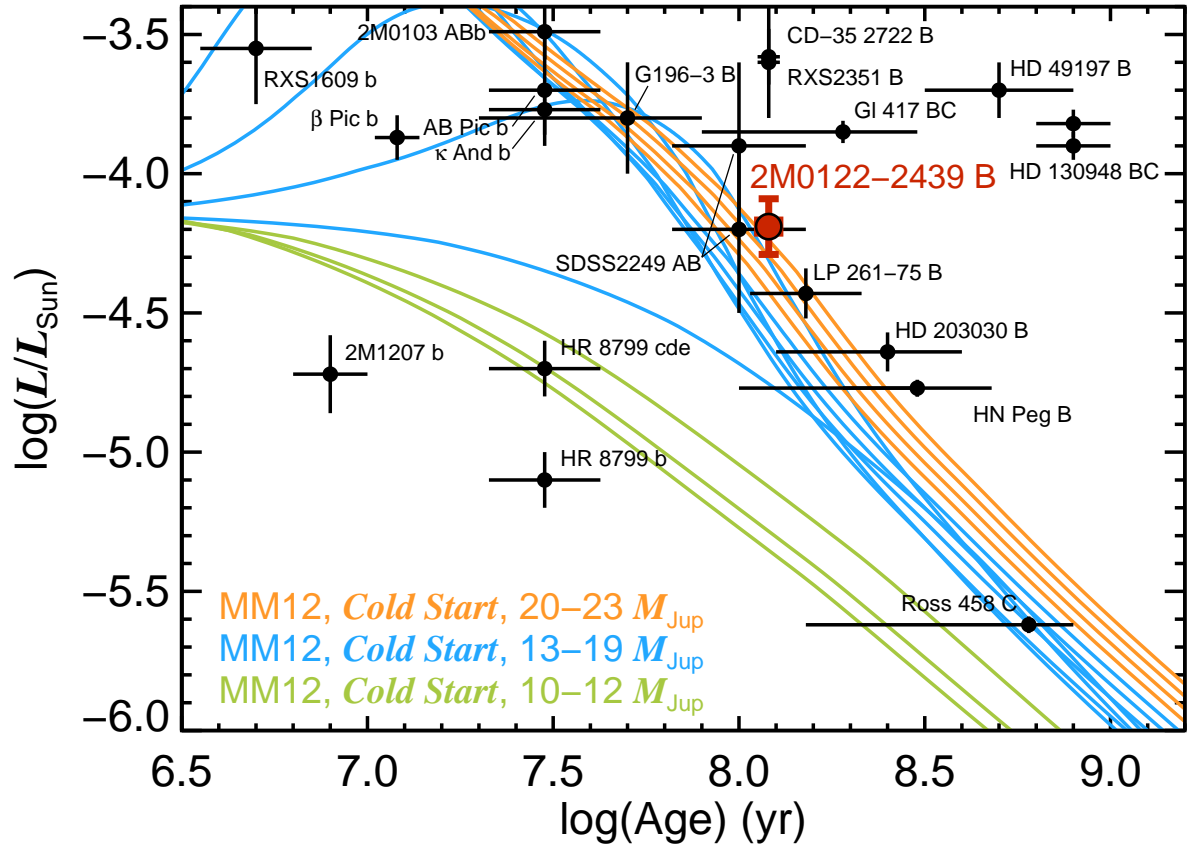}}
  \vskip -1.2in
  \caption{ The luminosity and age of 2MASS~0122--2439~B and other directly imaged companions
 compared with the cold start evolutionary models 
  from \citet{Molliere:2012p25183}.
  For masses above $\approx$13~\Mjup, deuterium burning temporarily increases the luminosity of objects formed 
  by core accretion.  Like the predictions from the hot start models (Figure~\ref{fig:hrd}), the cold start models 
  imply masses of $\approx$14~\Mjup \ and $\approx$23~\Mjup \ for 2MASS~0122--2439~B.
  Here the orange, blue, and green curves represent 10--12~\Mjup, 13--19~\Mjup, and 20--23~\Mjup \ cooling tracks
  in 1~\Mjup \ increments.
  See Figure~\ref{fig:hrd} for details about the other companions.
    \label{fig:coldhrd} } 
\end{figure}

\begin{figure}
  \hskip -1.7in
  \resizebox{10.5in}{!}{\includegraphics{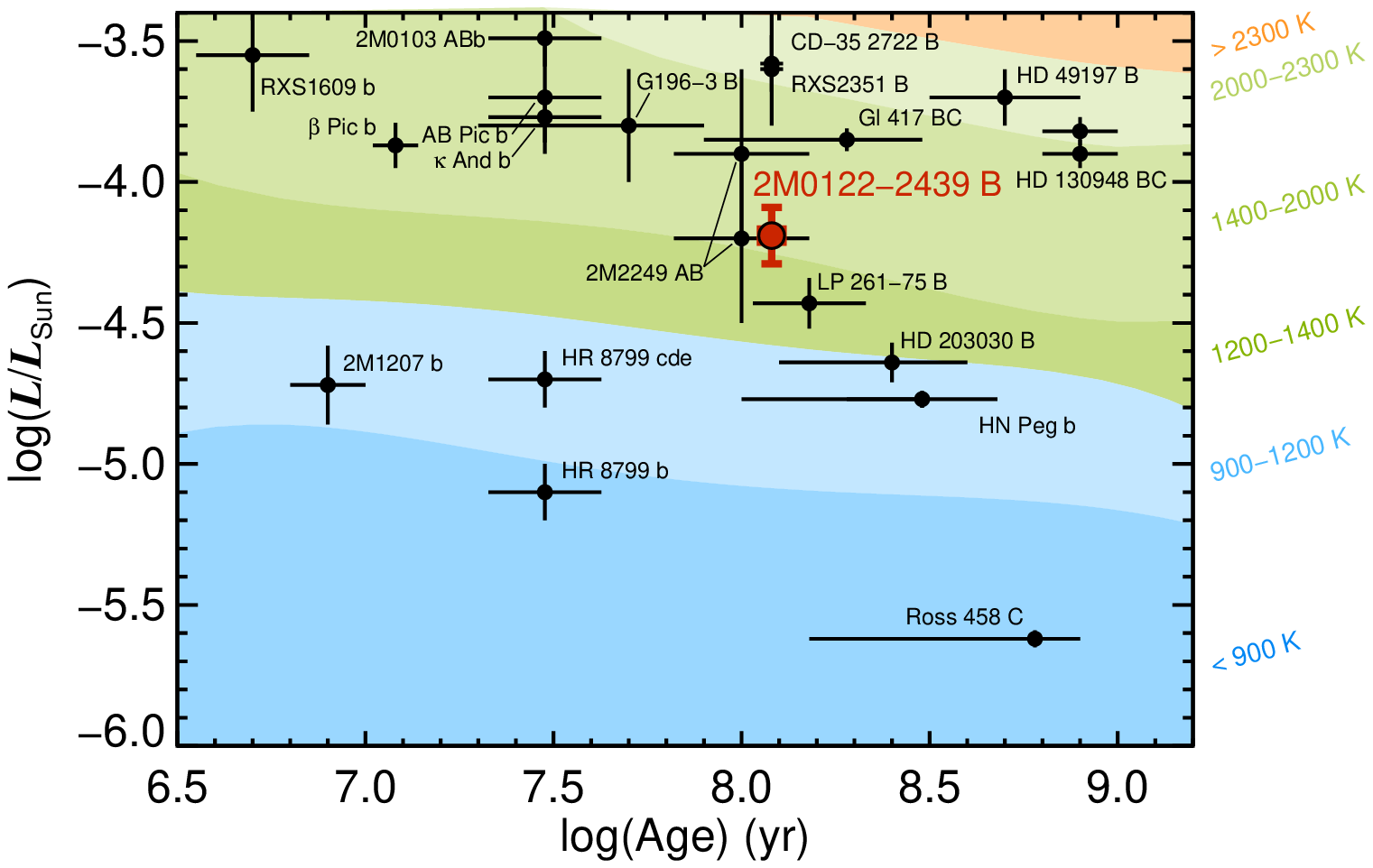}}
  \vskip -1.2in
  \caption{Effective temperature predictions from the \citet{Saumon:2008p14070} cooling models.
  Colors delineate temperature regimes corresponding to the following 
  (approximate) spectral classifications for field objects from \citet{Golimowski:2004p15703}: 
  $\le$M9 ($>$2300~K), L0--L3 (2000--2300~K), L3--L7 (1400--2000~K), L8--T2 (1200--1400~K), 
  T2--T7 (900--1200~K), $\ge$T8 ($<$900~K).
  Young substellar companions are overplotted for comparison (see Figure~\ref{fig:hrd} for details).
  The predicted effective temperature for 2MASS~0122--2439~B (1350--1500~K) corresponds to
  the L/T transition for field objects, but we observe a mid-L spectral type. 
 Like the HR~8799 planets, 2M1207--3932~b, and several other low-gravity L and T dwarfs,  2MASS~0122--2439~B 
 provides further evidence that the L/T transition occurs at lower temperatures for low surface gravities.
   \label{fig:twom0122_hrd_teff} } 
\end{figure}

The luminosity and age of 2MASS~0122--2439~B  
reside in a region where evolutionary models have dual-valued mass predictions.
Comparing other young ($<$1~Gyr) substellar
companions to the cooling models in Figure~\ref{fig:hrd} shows that  
several known objects also fall in or very close to this region: 
2MASS~J01033563--5515561~ABb (\citealt{Delorme:2013p25184}),
AB~Pic~b (\citealt{Chauvin:2005p19642}), 
$\kappa$~And~b (\citealt{Carson:2013p24396}), 
G196-3~B (\citealt{Rebolo:1998p19498}), 
SDSS~J224953.47+004404.6~B (\citealt{Allers:2010p20499}), 
LP261-75~B (\citealt{Reid:2006p22856}), HD 203030 B (\citealt{Metchev:2006p10342}),
and HN Peg B (\citealt{Luhman:2007p10341}).  
For these objects, a luminosity and age do not translate into a unique mass prediction,
instead being consistent with both $\approx$12--14~\Mjup \  and
$\approx$20--26~\Mjup \ mass tracks.

\begin{figure}
  \vskip -1in
  \hskip 0.6in
  \resizebox{7in}{!}{\includegraphics{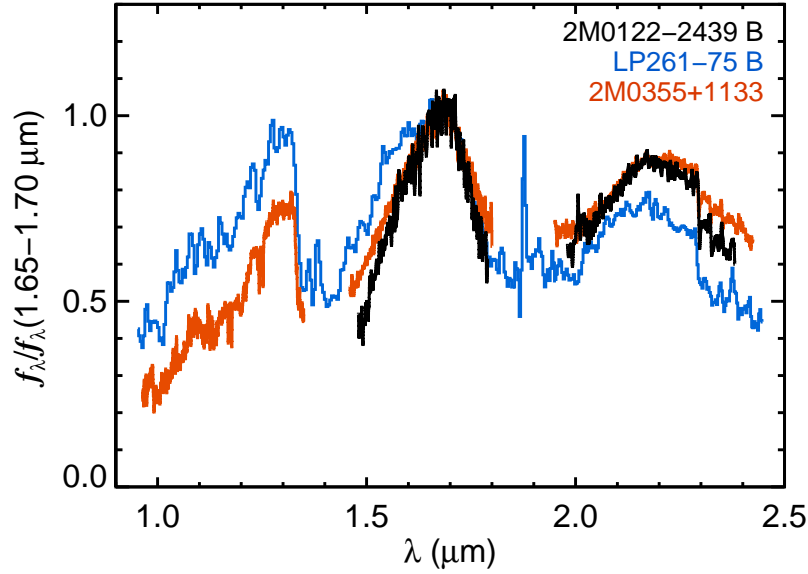}}
  \vskip -0.5in
  \caption{Comparison of 2MASS~0122--2439~B with the young companion LP~261-75~B  and 
  the L5 AB~Dor member 2MASS~0355+1133.
  2MASS~0122--2439~B and LP~261-75~B have similar ages and luminosities, but
  their spectra are very different.  The angular $H$-band shape of 
  2MASS~0122--2439~B indicates it has a lower surface gravity, pointing to a younger age and/or lower mass.
  The $H$-band shape is closer to (but slightly narrower than) that of  2MASS~0355+1133, but the 
  depth of the 2.3~$\mu$m CO feature is much greater in 2MASS~0122--2439~B. 
  \label{fig:lp261comp} } 
\end{figure}

One possible way to break this degeneracy may be by comparing the spectra of
objects with very similar luminosities and ages but different masses.  A lower
mass object will have a lower surface gravity which would be reflected in 
gravity-sensitive features in its spectrum.
At the age and luminosity 2MASS~0122--2439~B, the \citet{Saumon:2008p14070} evolutionary models predict
differences of $\approx$0.36~dex in log~$g$ between the lower (deuterium-burning) and higher 
(post-deuterium-burning) mass regimes.
LP~261-75~B has a similar age and luminosity,
but our spectra of the two companions are substantially different from each other (Figure~\ref{fig:lp261comp}).  
The more angular $H$ band of 2MASS~0122--2439~B suggests a lower gravity than LP~261-75~B, perhaps  
indicating that 2MASS~0122--2439~B belongs to the lower-mass ($\approx$13~\Mjup) set of model tracks while
LP~261-75~B belongs to the higher-mass ($\approx$25~\Mjup) set.  However, uncertainties in the ages of
these systems are large, so this might instead point to
a younger relative age for 2MASS~0122--2439~B.
SDSS~2244+0044~B also has a similar age and luminosity as 2MASS~0122--2439~B 
and may also be burning deuterium (\citealt{Allers:2010p20499}), 
but resolved spectroscopy of this binary
has only been obtained in $K$ band, which is less sensitive to differences in gravity than $J$ and $H$ bands 
so we cannot use this as a comparison point.
Similarly, while the absolute magnitude and colors of 2MASS~0122--2439~B are consistent with 
2MASS~0355+1133, the $H$-band shape is somewhat narrower and the 2.3~$\mu$m CO absorption is significantly
stronger in 2MASS~0122--2439~B.
As discussed in \citet{Allers:2013p25081}, this bolsters the notion there may be a diversity of 
spectral shapes at a 
given luminosity and age among young dusty L dwarfs; 
more discoveries are needed to map 
the influence of gravity, clouds, and temperature in this regime.

\begin{figure}
  \vskip -2.in
  \hskip .3in
  \resizebox{7in}{!}{\includegraphics{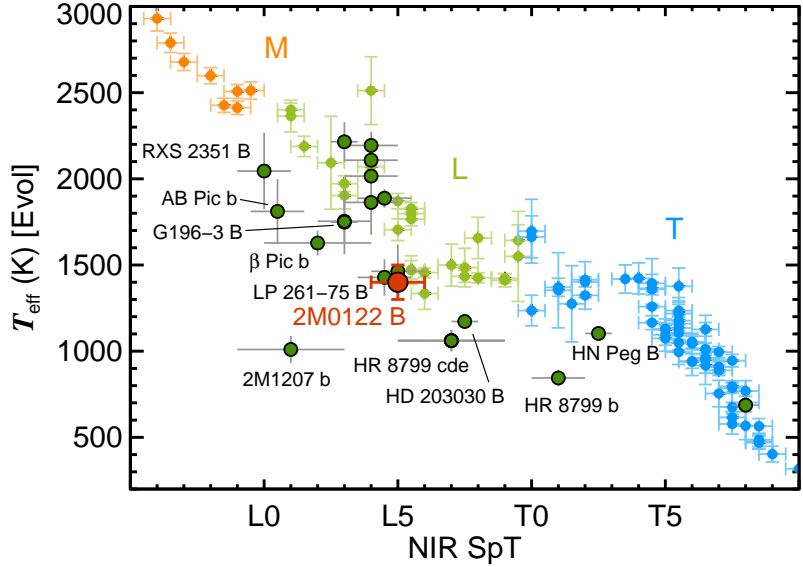}}
  \vskip -.7in
  \caption{Near-infrared spectral types of old ($>$1~Gyr) field objects and young ($<$1~Gyr) companions (green) compared
  to their predicted effective temperatures from the \citet{Burrows:1997p2706}
evolutionary models based on their ages and luminosities.  Field objects are from the \citet{Dupuy:2012p23924} compilation
 and include only single objects with normal spectral properties and 
 measured parallaxes.  Luminosities are computed using the $H$-band bolometric correction from \citet{Liu:2010p21195}, 
 and ages of 5~$\pm$~2~Gyr are assumed when deriving temperatures based on an interpolated grid of evolutionary models.  
 Ages and luminosities of the young companions are the same as
in Figure~\ref{fig:hrd}, except here we only include objects with classifications based on near-infrared spectroscopy.  
Note that for some objects (e.g., 2MASS~1207--3932~b) we have made use of updated near-infrared 
spectral types from \citet{Allers:2013p25081}.   Young companions with spectral types offset from the field
population are labeled; the remaining objects (in order of decreasing spectral type) are CD--35~2722~B (L3), 
SDSS~2249+0044~A (L3), 1RXS~160929.1--210524~b (L4), HD~49197~B (L4), HD~130958~BC (L4), 
Gl~417~BC (L4.5), SDSS~2249+0044~B (L5), 
and Ross~458~C (T8).
        \label{fig:teffspt} } 
\end{figure}

For field objects, 
the effective temperature for 2MASS~0122--2439~B from evolutionary models 
(1350--1500~K, Section~\ref{sec:lum})
corresponds to spectral types spanning the
L/T transition ($\sim$L7--T2). However, we find that 2MASS~0122--2439~B  better matches young mid-L spectral types.  
A similar disagreement between evolutionary model predictions and the 
temperatures estimated from spectral classification has been noted for a handful of
low-temperature substellar objects: HD~203030~B (\citealt{Metchev:2006p10342}),
HN~Peg~B (\citealt{Luhman:2007p10341}), 
2MASS~1207--3932~b (\citealt{Skemer:2011p22100}; \citealt{Barman:2011p22429}),
and the HR~8799 planets (\citealt{Bowler:2010p21344}; \citealt{Currie:2011p21928}; \citealt{Barman:2011p22429}).
This phenomenon is illustrated in Figure~\ref{fig:teffspt}, which shows the observed NIR spectral types of old field objects 
and young companions compared to the temperatures predicted by the \citet{Burrows:1997p2706}
evolutionary models based on their ages and luminosities.  
For a given predicted effective temperature, low gravity objects tend to have earlier spectral types than the field population.
This suggests that the spectral type-effective temperature sequence is also a function of gravity (age),  
becoming most apparent 
for young objects near the L/T transition 
when dust begins to settle below the photosphere at a fixed effective temperature of $\approx$1200--1400~K (\citealt{Saumon:2008p14070}).
Interestingly, several other young companions (1RXS~2351+3127~B, AB~Pic~b, G196-3~B, and $\beta$~Pic~b) 
may also exhibit this phenomenon at \emph{higher} effective temperatures, but more precise 
spectral types are needed to confirm this.
For 2MASS~0122--2439~B, a parallax measurement together with broader NIR spectroscopic coverage 
will enable a more detailed analysis of the gravity-dependent L/T transition.

  \acknowledgments

We thank Fred Vrba and the USNO group for sharing their parallax to LP~261-75~B;
Katelyn Allers, Zahed Wahhaj, Davy Kirkpatrick, David Lafreni\`{e}re, and Jennifer
Patience for the low gravity spectra used in this work;
Yosuke Minowa for assistance with the IRCS observations;  
Eric Nielsen for his compilation of YMG members;
and Katelyn Allers for assistance with the LP~261-75~B observations.
It is a pleasure to thank the Keck support astronomers and operating assistants
who helped make this work possible:
Joel Aycock, Randy Campbell, Marc Kassis, Jim Lyke, Terry Stickel, and Hien Tran.
B.P.B. and M.C.L. have been supported by NASA grant NNX11AC31G and NSF grant AST09-09222.
E.S. has been supported by NASA/GALEX grant NNX12AC18G.
We utilized data products from the Two Micron All Sky Survey, which is a joint project of the University of Massachusetts 
and the Infrared Processing and Analysis Center/California Institute of Technology, funded by the National Aeronautics and 
Space Administration and the National Science Foundation.
This publication makes use of data products from the Wide-field
    Infrared Survey Explorer, which is a joint project of the University of
    California, Los Angeles, and the Jet Propulsion Laboratory/California
    Institute of Technology, funded by the National Aeronautics and Space
    Administration (NASA).
 NASA's Astrophysics Data System Bibliographic Services, the VizieR catalogue access tool, and the SIMBAD database 
operated at CDS, Strasbourg, France, were invaluable resources for this work.
Finally, mahalo nui loa to the kama`\={a}ina of Hawai`i for their support of Keck and the Mauna Kea observatories.
We are grateful to conduct observations from this mountain.

\newpage


\clearpage

\begin{deluxetable}{lccccccc}
\tabletypesize{\scriptsize}
\tablewidth{0pt}
\tablecolumns{8}
\tablecaption{Adaptive Optics Imaging of 2MASS~J01225093--2439505~AB\label{tab:astrometry}}
\tablehead{
        \colhead{Date}   &   \colhead{Tel/Inst}  &  \colhead{Filter}  &  \colhead{$N$ $\times$ Coadds $\times$}   &  \colhead{FWHM}    &
        \colhead{Separation}    &    \colhead{PA}   &   \colhead{$\Delta$mag} \\
        \colhead{(UT)}   &   \colhead{}            &  \colhead{}      &    \colhead{Exp. Time (s)}    &  \colhead{(mas)}   &  
        \colhead{(")}    &    \colhead{($^{\circ}$)}   &   \colhead{}
        }   
\startdata
 2012 Oct 12    &    Subaru/IRCS   &  $K$  &  1~$\times$~5~$\times$1.0        &  128                        &   1.444~$\pm$~0.007      &   216.7~$\pm$~0.2      &    4.82~$\pm$~0.24     \\
 2013 Jan 18   &    Keck-II/NIRC2 &  $K$ &  5~$\times$~50~$\times$~0.3  &  55.1~$\pm$~1.4     &   1.4486~$\pm$~0.0006  &  216.64~$\pm$~0.08   &  5.36 $\pm$ 0.04  \\
 2013 Jan 18   &    Keck-II/NIRC2 &  $H$ &  3~$\times$~50~$\times$~0.3  &  51~$\pm$~2           &   1.4495~$\pm$~0.0015  &  216.59~$\pm$~0.08   &  6.18 $\pm$ 0.04  \\
 2013 Jan 18   &    Keck-II/NIRC2 &  $J$ &  3~$\times$~50~$\times$~0.3  &  50~$\pm$~4            &   $\cdots$                            &  $\cdots$   &  $>$5.8  \\
 2013 Jan 19   &    Keck-II/NIRC2 &  $L'$ &  4~$\times$~100~$\times$~0.5  &  101~$\pm$~7      &   1.452~$\pm$~0.005              &  216.6~$\pm$~0.4   &   4.19 $\pm$ 0.03  \\
 2013 Jun 30   &    Keck-II/NIRC2 &  $K$ &  4~$\times$~20~$\times$~0.3  &  80~$\pm$~14     &   1.448~$\pm$~0.004  &  216.47~$\pm$~0.07   &  5.35 $\pm$ 0.04  \\
 2013 Jun 30   &    Keck-II/NIRC2 &  $J$ &  5~$\times$~10~$\times$~2.0  &  120~$\pm$~20     &   1.433~$\pm$~0.010  &  216.9~$\pm$~0.4   &  6.79 $\pm$ 0.14  
\enddata
\tablecomments{NIRC2 FWHM measurements are computed using the IDL routine \texttt{NIRC2STREHL} made available by Keck Observatory.  The IRCS
FWHM measurement is for a single image.}
\end{deluxetable}

\begin{deluxetable}{lcccc}
\tabletypesize{\scriptsize}
\tablewidth{0pt}
\tablecolumns{5}
\tablecaption{Spectroscopy of 2MASS~J01225093--2439505~AB\label{tab:spectroscopy}}
\tablehead{
        \colhead{Date}   &   \colhead{Target}   &  \colhead{Telescope/}  &  \colhead{Filter}  &  \colhead{$N$ $\times$ Exp. Time}     \\
       \colhead{(UT)}   &     \colhead{}               & \colhead{Instrument}            &  \colhead{}      &    \colhead{(s)}     
        }   
\startdata
 2013 Feb 02    &  2MASS~0122--2439 B   &  Keck-I/OSIRIS  &  $Kbb$  &  7~$\times$~300      \\
 2013 Feb 02    &  2MASS~0122--2439 B   &  Keck-I/OSIRIS  &  $Hbb$  &  6~$\times$~300      \\
 2012 Dec 28    &  2MASS~0122--2439 A   &  Keck-I/HIRES  &  GG475  &  1~$\times$~90      
\enddata
\end{deluxetable}

\begin{deluxetable}{lcc}
\tabletypesize{\scriptsize}
\tablewidth{0pt}
\tablecolumns{3}
\tablecaption{Properties of the 2MASS~J01225093--2439505~AB System\label{tab:properties}}
\tablehead{
        \colhead{Parameter}   &    \colhead{Primary}    &    \colhead{Secondary}
        }   
\startdata
\multicolumn{3}{c}{Physical Properties} \\
\tableline
Age (Myr)                                                         &     \multicolumn{2}{c}{120~$\pm$~10\tablenotemark{a}}                               \\
$d_{phot}$ (pc)                                              &     36 $\pm$ 4\tablenotemark{b}                                               &  $\cdots$   \\
$\mu_{\alpha}$cos$\delta$ (mas/yr)          & 120.2 $\pm$ 1.9\tablenotemark{c}                                           &  $\cdots$    \\
$\mu_{\delta}$ (mas/yr)                                & --120.3 $\pm$ 1.7\tablenotemark{c}                                         &  $\cdots$    \\
$RV$ (km/s)                                    &   9.6 $\pm$~0.7                                           &  $\cdots$    \\
Proj. Sep. (AU)                                               &     \multicolumn{2}{c}{52 $\pm$ 6}                                                                     \\
$U$ (km/s)    &   \multicolumn{2}{c}{--5.4 $\pm$ 0.6}   \\
$V$ (km/s)    &   \multicolumn{2}{c}{--29 $\pm$ 3}   \\
$W$ (km/s)    &   \multicolumn{2}{c}{--8.0 $\pm$ 0.7}   \\
$X$ (pc)    &   \multicolumn{2}{c}{--4.5 $\pm$ 0.5}   \\
$Y$ (pc)    &   \multicolumn{2}{c}{--1.25 $\pm$ 0.14}   \\
$Z$ (pc)    &   \multicolumn{2}{c}{--36 $\pm$ 4}   \\
log($L_\mathrm{X}$/$L_\mathrm{Bol}$)    &     \multicolumn{2}{c}{--3.17~$\pm$~0.3}   \\
log($L_\mathrm{Bol}$/$L_{\odot}$)    &     --1.72~$\pm$~0.11   &  --4.19~$\pm$~0.10   \\
Spectral Type     &     M3.5~$\pm$~0.5\tablenotemark{d}   & L4--L6   \\
Mass     &     0.40~$\pm$~0.05~\Msun\tablenotemark{a}   &  12--14 or 23--27~\Mjup\tablenotemark{a, e}   \\
\tableline
\multicolumn{3}{c}{Photometry} \\
\tableline

$V$ (mag)                                                    &     14.24 $\pm$ 0.07\tablenotemark{c}   &  $\cdots$   \\
$R$ (mag)                                                    &     13.2\tablenotemark{f}   &  $\cdots$   \\
$I$ (mag)                                                      &     11.3\tablenotemark{f}    &  $\cdots$   \\
$J_\mathrm{MKO}$ (mag)     &     10.02 $\pm$ 0.03\tablenotemark{g} &  16.81 $\pm$ 0.14  \\
$H_\mathrm{MKO}$ (mag)     &     9.47 $\pm$ 0.02\tablenotemark{g}  &  15.65 $\pm$ 0.04 \\
$K_\mathrm{MKO}$ (mag)     &     9.17 $\pm$ 0.03\tablenotemark{g}  &  14.53 $\pm$ 0.05   \\
$L'$ (mag)                                 &     9.0 $\pm$ 0.1\tablenotemark{h}  &  13.2 $\pm$ 0.1   \\
$M_J$ (mag)\tablenotemark{i}                            &      7.2 $\pm$ 0.2   & 14.0 $\pm$ 0.3 \\
$M_H$ (mag)\tablenotemark{i}                            &      6.7 $\pm$ 0.2   & 12.9 $\pm$ 0.2 \\
$M_K$ (mag)\tablenotemark{i}                            &       6.4 $\pm$ 0.2   &  11.7 $\pm$  0.3 \\
$M_{L'}$ (mag)\tablenotemark{i}                          &      6.2 $\pm$ 0.3    &   10.4 $\pm$  0.3 \\
($J$--$H)_\mathrm{MKO}$ (mag)  &  0.55 $\pm$ 0.04   &  1.16 $\pm$ 0.15  \\
($H$--$K)_\mathrm{MKO}$ (mag)  &  0.30 $\pm$ 0.04   &  1.12 $\pm$ 0.06  \\
($J$--$K)_\mathrm{MKO}$ (mag)  &  0.85 $\pm$ 0.04   &  2.28 $\pm$ 0.15  \\
$K_\mathrm{MKO}$--$L'$ (mag)                               &  0.30 $\pm$ 0.10   &  1.33 $\pm$ 0.11  \\
$WISE$ $W1$ (mag)            &      9.01 $\pm$ 0.03\tablenotemark{j}  &  $\cdots$   \\
$WISE$ $W2$ (mag)            &     8.84 $\pm$ 0.02\tablenotemark{j}  &  $\cdots$  \\
$WISE$ $W3$ (mag)            &      8.75 $\pm$ 0.03\tablenotemark{j}  &  $\cdots$  \\
$WISE$ $W4$ (mag)            &     8.2 $\pm$ 0.2\tablenotemark{j}  &  $\cdots$  \\
$GALEX$ $NUV$ (mag)    &     20.67 $\pm$ 0.14\tablenotemark{k}    &  $\cdots$   \\
$GALEX$ $FUV$ (mag)    &     21.2 $\pm$ 0.3\tablenotemark{k}    &  $\cdots$   \\
$ROSAT$ flux (erg sec$^{-1}$ cm$^{-2}$)    &     3.4~$\pm$~1.6$\times$10$^{-13}$    &  $\cdots$   
\enddata
\tablecomments{$UVWXYZ$ values and absolute magnitudes are based on the photometric distance estimate.  
$U$ and $X$ are positive toward the
Galactic center, $V$ and $Y$ are positive toward the direction of galactic rotation, and $W$ and $Z$ are positive toward the
North Galactic Pole.}
\tablenotetext{a}{Assumes membership to AB Dor YMG (120~Myr; \citealt{Luhman:2005p22437}; \citealt{Barenfeld:2013p24822}), 
otherwise a wider range of 10--120~Myr is possible.}
\tablenotetext{b}{Computed using the Pleiades $M_V$ vs. $V-K$ relation from Section~\ref{sec:distance}.}
\tablenotetext{c}{From the UCAC4 catalog (\citealt{Zacharias:2013p24823}).  Systematic errors in proper motions 
are estimated to be $\sim$1--4~mas yr$^{-1}$.}
\tablenotetext{d}{From \citet{Riaz:2006p20030}.}
\tablenotetext{e}{Hot start evolutionary models overlap near the deuterium-burning limit, resulting in dual-valued mass predictions (Section~\ref{sec:lum}).}
\tablenotetext{f}{From the USNO-B1.0 catalog (\citealt{Monet:2003p17612}).}
  \tablenotetext{g}{Converted to the MKO system from 2MASS (\citealt{Skrutskie:2006p589}) using 
  the transformation in \citet{Leggett:2006p2674}.}
\tablenotetext{h}{Assumes a $K$--$L'$ color of 0.2~$\pm$~0.1~mag (\citealt{Golimowski:2004p15703}).}
\tablenotetext{i}{Based on our photometric distance estimate.}
\tablenotetext{j}{From the $WISE$ All-Sky Data Release (\citealt{Cutri:2012p25019}).}
\tablenotetext{k}{From the $GALEX$ GR6/7 data release (\citealt{Morrissey:2007p22251}).}
\end{deluxetable}

\clearpage

\clearpage

\end{document}